\documentstyle[11pt,amssymb,epsf]{article}

\def\cramper{\medmuskip = 2mu plus 1mu minus 1mu}

\def\ben{\begin{equation}}
\def\een{\end{equation}}

\let\a=\alpha    
    
  \let\n=\nu   
\let\s=\sigma    \let\c=\chi 
       \let\D=\Delta  
     
\let\C=\Chi

\def\nn{\nonumber} \def\bd{\begin{document}} \def\ed{\end{document}}
\def\ds{\documentstyle} \let\fr=\frac \let\bl=\bigl \let\br=\bigr
\let\Br=\Bigr \let\Bl=\Bigl
\let\bm=\bibitem
\let\na=\nabla
\let\pa=\partial \let\ov=\overline
\newcommand{\be}{\begin{equation}}
\newcommand{\ee}{\end{equation}}
\def\ba{\begin{array}}
\def\ea{\end{array}}
\def\ft#1#2{{\textstyle{{\scriptstyle #1}\over {\scriptstyle #2}}}}
\def\fft#1#2{{#1 \over #2}}
\def\del{\partial}
\def\vp{\varphi}
\def\sst#1{{\scriptscriptstyle #1}}
\def\oneone{\rlap 1\mkern4mu{\rm l}}
\def\td{\tilde}
\def\wtd{\widetilde}
\def\ie{\rm i.e.\ }
\def\dalemb#1#2{{\vbox{\hrule height .#2pt
        \hbox{\vrule width.#2pt height#1pt \kern#1pt
                \vrule width.#2pt}
        \hrule height.#2pt}}}
\def\square{\mathord{\dalemb{6.8}{7}\hbox{\hskip1pt}}}
\newcommand{\ho}[1]{$\, ^{#1}$}
\newcommand{\hoch}[1]{$\, ^{#1}$}
\newcommand{\bea}{\begin{eqnarray}}
\newcommand{\eea}{\end{eqnarray}}
\newcommand{\ra}{\rightarrow}
\newcommand{\lra}{\longrightarrow}
\newcommand{\Lra}{\Leftrightarrow}
\newcommand{\bp}{\tilde \beta^\prime}
\newcommand{\tr}{{\rm tr} }
\newcommand{\Tr}{{\rm Tr} }
\def\0{{\sst{(0)}}}
\def\1{{\sst{(1)}}}
\def\2{{\sst{(2)}}}
\def\3{{\sst{(3)}}}
\def\4{{\sst{(4)}}}
\def\5{{\sst{(5)}}}
\def\6{{\sst{(6)}}}
\def\7{{\sst{(7)}}}
\def\8{{\sst{(8)}}}
\def\n{{\sst{(n)}}}
\def\cA{{{\cal A}}}
\def\cB{{{\cal B}}}
\def\cF{{{\cal F}}}
\def\cH{{{\cal H}}}
\def\tV{\widetilde V}
\def\tW{\widetilde W}
\def\tH{\widetilde H}
\def\tE{\widetilde E}
\def\tF{\widetilde F}
\def\tA{\widetilde A}
\def\im{{i}}
\def\tY{{{\wtd Y}}}
\def\ep{{\epsilon}}
\def\vep{{\varepsilon}}
\def\R{\rlap{\rm I}\mkern3mu{\rm R}}
\def\bD{{{\bar D}}}

\def\R{\rlap{\rm I}\mkern3mu{\rm R}}
\def\bD{{{\bar D}}}
\def\R{{{\Bbb R}}}
\def\C{{{\Bbb C}}}
\def\H{{{\Bbb H}}}
\def\CP{{{\Bbb C}{\Bbb P}}}
\def\RP{{{\Bbb R}{\Bbb P}}}
\def\Z{{{\Bbb Z}}}
\def\bA{{{\Bbb A}}}
\def\bB{{{\Bbb B}}}
\def\bC{{{\Bbb C}}}
\def\bD{{{\Bbb D}}}
\def\bE{{{\Bbb E}}}
\def\bZ{{{\Bbb Z}}}
\def\Re{{{\frak{Re}}}}
\def\Im{{{\frak{Im}}}}
\def\cosec{{\,\hbox{cosec}\,}}
\def\Gm{{\Gamma_{\!\! -}}}
\def\Gp{{\Gamma_{\!\! +}}}
\def\stan{{standard }}
\def\nonstan{{supernumerary }}

\newcommand{\tamphys}{\it Center for Theoretical Physics,
Texas A\&M University, College Station, TX 77843}

\newcommand{\upenn}{\it Department of Physics and Astronomy,\\ University
of Pennsylvania, Philadelphia, PA 19104}

\newcommand{\brussels}{\it Physique Th\'eorique et Math\'ematique,
Universit\'e Libre de Bruxelles,\\ Campus Plaine C.P. 231, B-1050
Bruxelles, Belgium}

\newcommand{\auth}{Z.-W. Chong\hoch{\ddagger1}, M. Cveti\v c\hoch{*2},  
H. L\"u\hoch{\ddagger1} and C.N. Pope\hoch{\ddagger1}}

\begin{document}

\begin{flushright}
MIFP-04-20\ \ \ UPR-1096-T \\
{\bf hep-th/0411045}\\
November\  2004
\end{flushright}

\vspace{10pt}

\begin{center}

{\large {\bf  Charged Rotating Black Holes in 
Four-Dimensional Gauged and Ungauged Supergravities}}

\vspace{20pt}
\auth

\vspace{10pt}{\hoch{\dagger}\it George P. \& Cynthia W. Mitchell
Institute for Fundamental Physics,\\ Texas A\& M University,
College Station, TX 77843-4242, USA}

\vspace{10pt}{\hoch{*}\it Department of Physics and Astronomy,\\
University of Pennsylvania, Philadelphia, PA 19104, USA}


%
%
%

\vspace{20pt}


\begin{abstract}

    We study four-dimensional non-extremal charged rotating black
holes in ungauged and gauged supergravity.  In the ungauged case, we
obtain rotating black holes with four independent charges, as
solutions of ${\cal N}=2$ supergravity coupled to three abelian vector
multiplets.  This is done by reducing the theory along the time
direction to three dimensions, where it has an $O(4,4)$ global
symmetry.  Applied to the reduction of the uncharged Kerr metric,
$O(1,1)^4\subset O(4,4)$ transformations generate new solutions that
correspond, after lifting back to four dimensions, to the introduction
of four independent electromagnetic charges.  In the case where these
charges are set pairwise equal, we then generalise the
four-dimensional rotating black holes to solutions of gauged ${\cal
N}=4$ supergravity, with mass, angular momentum and two independent
electromagnetic charges.  The dilaton and axion fields are
non-constant.  We also find generalisations of the gauged and ungauged
solutions to include the NUT parameter, and for the ungauged
solutions, the acceleration parameter too.  The solutions in gauged
supergravity provide new gravitational backgrounds for a further study
of the AdS$_4$/CFT$_3$ correspondence at non-zero temperature.

\end{abstract}
\end{center}

{\vfill\leftline{}\vfill \vskip 10pt \footnoterule {\footnotesize
{\footnotesize
\hoch{1} Research supported in part by DOE grant
DE-FG03-95ER40917.}\vskip 2pt
\hoch{2} Research supported in part by DOE grant
DE-FG02-95ER40893, NSF grant INTO3-24081, and the\\
$\phantom{xxxxi}$  Fay R. and Eugene L.
Langberg Chair.}\vskip 2pt
}

\pagebreak

\tableofcontents
\addtocontents{toc}{\protect\setcounter{tocdepth}{2}}
\newpage

\section{Introduction}

     Rotating charged black hole solutions of ungauged supergravity
play an important role in the microscopic study of black hole entropy.
It turns out that the microscopic properties can be addressed
quantitatively not only for BPS black holes, but also for black holes
that are close to extremality.  (For a recent review see \cite{Mathur},
and references therein.)  The prerequisite for these studies is to obtain
explicit black hole solutions on the supergravity side. These solutions
are typically characterised by multiple electromagnetic charges, in
addition to the mass and angular momenta.  In five dimensions, such
explicit non-extremal solutions, specified by their mass, two angular
momentum parameters and three charges, were found in \cite{CY5}. They
were obtained by employing a solution-generating technique using global
symmetries of a four-dimensional theory that was obtained by reducing
from five dimensions along the time direction.  The first examples
of electrically-charged rotating black holes in ungauged supergravity
theories were obtained, by using such a solution-generating technique,
in \cite{Sen}.

    Employing analogous methods, the explicit metric and scalar
fields for four-dimensional four-charge rotating black holes were
obtained in \cite{CY4}.  Unfortunately, the explicit form of the four
gauge potentials for this solution was not given explicitly in \cite{CY4}.
Further work, again using the same solution-generating procedure, produced
black holes in $4\le D\le 9$ dimensions parameterised by their mass, two
charges, and their $[(D-1)/2]$ angular momentum parameters \cite{CYD}.
These solutions of ungauged supergravity all provide gravitational
backgrounds for the microscopic study of black hole entropy within the
string theory framework.

   By contrast, black holes in gauged supergravity provide
gravitational backgrounds that are relevant to the AdS/CFT correspondence.
In particular, such non-extremal solutions play an important role in the
study of the dual field theory at non-zero temperature.  (An early study
of the implications of static charged AdS black holes \cite{Sabraetal} in
the dual theory was carried out in \cite{CGI,CGII}. For recent related
work see \cite{Buchel,SabraII,gubheck} and references therein.)  However,
the explicit form of charged AdS black hole solutions that are also
rotating has remained elusive until recently.  In \cite{CLPI,CLPII}
the first examples of non-extremal rotating charged AdS black holes
in five-dimensional ${\cal N}=4$ gauged supergravity were obtained,
in the special case where the two angular momenta $J_i$ are set equal.
These solutions are characterised by their mass, three electromagnetic
charges, and the angular momentum parameter $J=J_1=J_2$.  By taking
appropriate limits, one obtains the various supersymmetric charged
rotating $D=5$ black holes obtained in \cite{gutreal1,gutreal2,klesab}.
If instead the charges are set to zero, the solutions reduce to
the rotating AdS$_5$ black hole constructed in \cite{hhtr}, with
$J_1=J_2$.

    In four dimensions, there should exist rotating black hole solutions
in gauged ${\cal N}=8$ supergravity with four independent electromagnetic
charges.  Until now, the only known solutions of this type were the
Kerr-Newman-AdS black holes \cite{carter1,carter2}, which correspond to
setting the four electromagnetic charges equal.

    One of the purposes of this paper is first to construct the
complete and explicit form of the general rotating black holes of
four-dimensional {\it ungauged} supergravity, with four independent
electromagnetic charges.  They can be viewed as solutions in ungauged
${\cal N}=2$ supergravity coupled to three vector multiplets, which in
turn can be embedded in ${\cal N}=8$ maximal supergravity.  We employ
a solution-generating technique in which the ${\cal N}=2$ theory is
reduced to three dimensions on the time direction, where it has an
$O(4,4)$ global symmetry.  By acting on the reduction of the uncharged
Kerr solution with an $O(1,1)^4\subset O(4,4)$ subgroup of the global
symmetry, we obtain a new solution that lifts back to a solution of the
four-dimensional theory with four independent electromagnetic charges.
In this formulation, two of the $U(1)$ charges are electric and two
are magnetic.  By obtaining the explicit form of the four $U(1)$ gauge
potentials, as well as the other fields, we therefore complete the results
in \cite{CY4}, where the metric and scalar fields were found.  We then
apply the same generating technique to generalise these results by the
inclusion also of the NUT parameter, and the acceleration parameter.

   The second goal of this paper is to obtain charged rotating black
hole solutions in four-dimensional {\it gauged} supergravity.  We have
been able to do this in the case where the four charges of the
ungauged theory are first set pairwise equal.  With this restriction,
we are able to conjecture, and then explicitly verify, the expression
for the two-charge rotating black hole solutions of ${\cal N}=4$
gauged supergravity.  The solutions have varying dilaton and axion
fields.

   The paper is organised as follows.  In Section 2 we describe the
solution-generating technique for constructing charged rotating black
holes of four-dimensional ungauged supergravity.  In fact the
procedure can be used to introduce four independent charges in any
Ricci-flat four dimensional metric admitting a timelike Killing
vector.  In Section 3 we present the explicit form of the four-charge
rotating black hole solution, generated from the Kerr metric, and give
its specialisation to the case where the charges are set pairwise
equal.  In Section 4 we present the generalisation of this latter case
to a solution in ${\cal N}=4$ gauged supergravity.  In Section 5,
generalisations of both the gauged and ungauged supergravity solutions
are found, in which in addition the NUT parameter is non-zero.  In
Section 6 we obtain a further generalisation of the ungauged
supergravity solutions to include also the acceleration parameter.  In
Appendix A we give a matrix realisation of the generators of $O(4,4)$,
which is helpful for implementing the explicit transformations in
three dimensions.  In Appendix B we give the explicit form of the
rotating four-charge solution with the NUT parameter for the ungauged
case.  Appendix C contains a discussion of the supersymmetry of
extremal rotating AdS black holes in four dimensions.  Concluding
remarks are given in Section 7.

\section{Charge-Generating Procedure}\label{cgensec}

    In this section, we set up the basic formalism for generating
four-dimensional configurations carrying 4 independent charges, that
are solutions of ungauged ${\cal N}=2$ supergravity coupled to three
vector multiplets. This theory, and hence also its solutions, can be
consistently embedded in four-dimensional ${\cal N}=8$ supergravity.
The procedure involves starting with an uncharged four-dimensional
solution that has a timelike Killing vector $\del/\del t$, and
reducing it to three dimensions on the $t$ direction.  The reduction
of the ${\cal N}=2$ theory itself yields a three-dimensional theory
with an $O(4,4)$ global symmetry, after all the three-dimensional
vector fields have been dualised to axions.  By acting with an
$O(1,1)^4$ subgroup of $O(4,4)$ on the dimensionally reduced solution,
we generate new solutions involving four parameters $\delta_i$
characterising the $O(1,1)^4$ transformation.  Upon undualising the
transformed dualised axions back to vectors again, and lifting back
to $D=4$, we thereby arrive at supergravity solutions carrying 4
electromagnetic charges, parameterised by the $\delta_i$.

   In this section, we shall present the three-dimensional results for
the reduction and $O(1,1)^4$ transformation of a general
four-dimensional uncharged solution with a timelike Killing vector
$\del/\del t$.  One cannot abstractly ``undualise'' the
three-dimensional scalars that originate from vectors in $D=4$ (or
from the Kaluza-Klein vector), since dualisation is intrinsically a
non-local procedure.  However, once one has an explicit solution, the
process of undualisation can be implemented explicitly.  Thus, in
subsequent sections we shall apply these results to particular cases,
and implement the complete and explicit construction of the charged
four-dimensional solutions.  (A solution-generating technique of the
type we are using here was first employed in \cite{Sen}, to obtain
electrically charged rotating black holes in ungauged supergravity.)
As we shall see later, our solutions will carry two electric and two
magnetic charges.

\subsection{$O(4,4)$ symmetry of the reduced $D=3$ theory}\label{symsec}

   The four-dimensional Lagrangian for the bosonic sector of the
${\cal N}=2$ supergravity coupled to three vector multiplets can be
written as\footnote{Our conventions for dualisation are that a
$p$-form $\omega$ with components defined by $\omega=1/p!\,
\omega_{i_1\cdots i_p}\, dx^{i_1}\wedge \cdots \wedge dx^{i_p}$ has
dual ${*\omega}$ with components $({*\omega})_{i_i\cdots i_{D-p}} =
1/p!\, \ep_{i_1\cdots i_{D-p} j_i\cdots j_p}\, \omega^{j_1\cdots
j_p}$.}
\bea
{\cal L}_4 &=& R\, {*\oneone} - \ft12 {*d\varphi_i}\wedge d\varphi_i 
   - \ft12 e^{2\varphi_i}\, {*d\chi_i}\wedge d\chi_i - \ft12 e^{-\varphi_1}\,
\Big( e^{\varphi_2-\varphi_3}\, {*\hat F_{\2 1}}\wedge \hat F_{\2 1}\nn\\
&& + e^{\varphi_2+\varphi_3}\, {* \hat F_{\2 2}}\wedge \hat F_{\2 2}
   + e^{-\varphi_2 + \varphi_3}\, {*\hat \cF_\2^1 }\wedge \hat \cF_\2^1 + 
     e^{-\varphi_2 -\varphi_3}\, {*\hat\cF_\2^2}\wedge \hat \cF_\2^2\Big)\nn\\
&& - \chi_1\, (\hat F_{\2 1}\wedge \hat\cF_\2^1 + 
                  \hat F_{\2 2}\wedge \hat\cF_\2^2)\,,
\label{d4lag}
\eea
where the index $i$ labelling the dilatons $\varphi_i$ and axions $\chi_i$
ranges over $1\le i \le 3$.  The four field strengths can be written in 
terms of potentials as
\bea
\hat F_{\2 1} &=& d \hat A_{\1 1} - \chi_2\, d\hat\cA_\1^2\,,\nn\\
\hat F_{\2 2} &=& d\hat A_{\1 2} + \chi_2\, d\hat\cA_\1^1 - 
    \chi_3\, d \hat A_{\1 1} +
      \chi_2\, \chi_3\, d\hat \cA_\1^2\,,\nn\\
\hat \cF_\2^1 &=& d\hat \cA_\1^1 + \chi_3\, d\hat \cA_\1^2\,,\nn\\
\hat \cF_\2^2 &=& d\hat \cA_\1^2\,.
\eea
Note that we are placing hats on the four-dimensional field strength 
and gauge potentials, to distinguish them from the three-dimensional 
fields.

   The four-dimensional theory can be obtained from six-dimensions, by 
reducing the bosonic string action 
\be
{\cal L}_6 = R\, {*\oneone} - \ft12 e^{-\sqrt2 \phi}\, {*F_\3}\wedge F_\3
\ee
on $T^2$.  Thus the four-dimensional Lagrangian itself has an
$O(2,2)\sim SL(2,\R)\times SL(2,\R)$ global symmetry, which enlarges
at the level of the equations of motion to include a third $SL(2,\R)$
factor when electric/magnetic S-duality transformations are included.
We are going to reduce it one stage further, to $D=3$.  If left in its
raw form, the three-dimensional Lagrangian would have an $O(3,3)$
global symmetry.  However, if the 1-form potentials in $D=3$ are
dualised to axions (so that there are only dilatons and axions, plus
the metric, in $D=3$), then as is well known, the global symmetry will
be enhanced to $O(4,4)$.  The reduction from $D=4$ to $D=3$ will be
performed on the time coordinate.  This will imply that the coset
parameterised by the dilatons and axions will not be
$O(4,4)/(O(4)\times O(4))$, as would be the case for a spacelike
reduction, but instead $O(4,4)/O(4,\C)$.

    To proceed, we first reduce the fields in the 
Lagrangian (\ref{d4lag}), according to the standard Kaluza-Klein reduction
scheme adapted to the case of a timelike reduction.  Thus we write the
following reduction ans\"atze for the metric and for 1-form potentials:
\bea
d\hat s_4^2 &=& - e^{\varphi_4}\, (dt + \cB_\1)^2 + e^{-\varphi_4}\, 
                  ds_3^2 \,,\label{gred}\\
\hat A_\1 &=& A_\1 + A_\0\, (dt+ \cB_\1)\,.\label{ared}
\eea
The field strengths reduce according to the rule
\be
\hat F_\2 = F_ 2 + F_1\wedge (dt+ \cB_\1)\,.
\ee

   In order to abbreviate the description, we shall directly present the 
fully-dualised form of the three-dimensional Lagrangian that results
from reducing (\ref{d4lag}) according to this scheme, and then indicate 
afterwards how the three-dimensional fields are related to the 
four-dimensional ones.  We find that the fully-dualised three-dimensional 
Lagrangian can be written as
\bea
e^{-1}\, {\cal L}_3 &=& R - \ft12 (\del\varphi_i)^2  
 -\ft12 e^{2\varphi_1}\,  (\del\chi_1)^2 
 -\ft12 e^{2\varphi_2}\,  (\del\chi_2)^2
-\ft12 e^{3\varphi_1}\,  (\del\chi_3)^2\nn\\
&& -\ft12 e^{-2\varphi_4}\, (\del\chi_4 + \sigma_1\, \del\psi_1 +
 \sigma_2\, \del\psi_2 + \sigma_3\ \del\psi_3 + \sigma_4\, \del\psi_4)^2\nn\\
&& +\ft12 e^{-\varphi_1+\varphi_2-\varphi_3-\varphi_4}\, (\del\sigma_1 -
                            \chi_2\, \del\sigma_4)^2\nn\\
 && +\ft12 e^{-\varphi_1+\varphi_2+\varphi_3-\varphi_4}\, (\del\sigma_2 +
 \chi_2\, \del\sigma_3-\chi_3\, \del\sigma_1 + \chi_2\, \chi_3\, 
                        \del\sigma_4)^2\nn\\
&& +\ft12 e^{-\varphi_1-\varphi_2+\varphi_3-\varphi_4}\, (\del\sigma_3 +
                \chi_3\, \del\sigma_4)^2 
  +\ft12 e^{-\varphi_1-\varphi_2-\varphi_3-\varphi_4}\,(\del\sigma_4)^2\nn\\
&&  +\ft12 e^{\varphi_1-\varphi_2+\varphi_3-\varphi_4}\,
   (\del\psi_1 + \chi_3\, \del\psi_2 -\chi_1\, \del\sigma_3 -
    \chi_1\, \chi_3\, \del\sigma_4)^2\nn\\
&&  +\ft12 e^{\varphi_1-\varphi_2-\varphi_3-\varphi_4}\,
   (\del\psi_2 -\chi_1\, \del\sigma_4)^2 \nn\\
&&  +\ft12 e^{\varphi_1+\varphi_2-\varphi_3-\varphi_4}\,
  (\del\psi_3 -\chi_2\, \del\psi_2 -\chi_1\, \del\sigma_1 +
     \chi_1\, \chi_2\, \del\sigma_4)^2\nn\\
&& +\ft12 e^{\varphi_1+\varphi_2+\varphi_3-\varphi_4}\,
(\del\psi_4 + \chi_2\, \del\psi_1 -\chi_3\, \del\psi_3 - \chi_1\, \del\sigma_2
  + \chi_2\, \chi_3\, \del\psi_2 \nn\\
&&\qquad\qquad\qquad\qquad - \chi_1\, \chi_2\,\del\sigma_3
  + \chi_1\, \chi_3\, \del\sigma_1 - \chi_1\, \chi_2\, \chi_3\, 
   \del\sigma_4)^2\,,\label{d3lag}
\eea
where now the $i$ index ranges over $1\le i \le 4$.  Note that the
last eight terms have the non-standard sign for their kinetic terms,
in consequence of the timelike reduction, and the subsequent
dualisations in a Euclidean-signature metric.  The axion $\chi_4$
corresponds to the dual of the Kaluza-Klein vector $\cB_\1$ in
(\ref{gred}); the axions $\sigma_i$ correspond to the components
$A_\0$ (as in (\ref{ared})) of the reductions of the 4 four-dimensional
potentials, in the order $(A_{\1 1}, A_{\1 2}, \cA_\1^1, \cA_\1^2)$; and
the axions $\psi_i$ correspond to the dualisations of the three-dimensional 
1-forms $A_\1$ in (\ref{ared}), taken in the same order as the $\sigma_i$.

   In detail, the dualisations are performed as follows.\footnote{These
results are obtained by applying the standard procedure of
introducing the dual potential as a Lagrange muliplier for the
Bianchi identity of the original field strength.}  The field
strength $G_\2=d \cB_\1$ for the Kaluza-Klein 1-form is replaced by
\be
e^{2\varphi_4}\, {*G_\2} = d\chi_4 + \sigma_1\, d\psi_1 + 
    \sigma_2\, d\psi_2 +\sigma_3\, d\psi_3 +   \sigma_4\, d\psi_4\,.
\label{kkdual}
\ee
The four field strengths coming from the four field strengths in four
dimensions are replaced by
\bea
- e^{-\varphi_1 + \varphi_2 - \varphi_3 + \varphi_4}\, {* F_{\2 1}} 
&=& d\psi_1 + \chi_3\, d\psi_2 - \chi_1\, d\sigma_3 - \chi_1\,
\chi_3\, d\sigma_4\,,\nn\\
- e^{-\varphi_1 + \varphi_2 + \varphi_3 + \varphi_4}\, {* F_{\2 2}} 
&=& d\psi_2 -\chi_1\, d\sigma_4\,,\nn\\
- e^{-\varphi_1 - \varphi_2 + \varphi_3 + \varphi_4}\, {* \cF_{\2}^1} 
&=& d\psi_3 - \chi_2\, d\psi_2 - \chi_1\, d\sigma_1 + \chi_1\,
\chi_2\, d\sigma_4\,,\nn\\
- e^{-\varphi_1 - \varphi_2 - \varphi_3 + \varphi_4}\, {* \cF_{\2}^2} 
&=& d\psi_4 + \chi_2\, d\psi_1 - \chi_3\, d\psi_3 -
\chi_1\,d\sigma_2 + \chi_2\, \chi_3\, d\psi_2 \nn\\
&&- \chi_1\, \chi_2\, d\sigma_3 +
 \chi_1\, \chi_3\, d\sigma_1 - \chi_1\, \chi_2\, \chi_3\, d\sigma_4\,.
\eea

   The Lagrangian (\ref{d3lag}) can be re-expressed as
\be
{\cal L}_3= R\, {*\oneone} - \ft12 {*d\varphi_i}\wedge d\varphi_i 
   -\ft12 \sum_{\a=1}^{12}\, \eta_\a\, e^{\vec a_\a\cdot\vec\varphi}\, 
   {*F^\a}\wedge F^\a\,,
\ee
where each of the twelve 1-form field strengths $F^\a$ can be read off
by comparison with the twelve axion kinetic terms in (\ref{d3lag}).
Likewise, the corresponding dilaton vector $\vec a_\a$ can be read off
from the dilatonic prefactor of each axion kinetic term, and the
coefficients $\eta_\a=\pm1 $ can be read off from the signs of the
kinetic terms.  One can then easily see that, as in the constructions in
\cite{cjlp1}, we have
\be
dF^\a = \ft12 f^\a{}_{\beta\gamma}\, F^\beta\wedge F^\gamma\,.\label{dfrels}
\ee
Introducing generators $E_\a$ and defining ${\cal F}\equiv F^\a\, E_\a$, 
one can express (\ref{dfrels}) as $d{\cal F} = {\cal F}\wedge{\cal F}$, 
where 
\be
[E_\a, E_\beta] = f^\gamma{}_{\a\beta}\, E_\gamma\,.
\ee
As in the discussions in \cite{cjlp1}, the $E_\a$ are easily seen to be
the positive-root generators of the global symmetry group $O(4,4)$.  We 
also introduce the four Cartan generators $\vec H$, which satisfy
\be
[\vec H, E_\a] = \vec a_\a\, E_\a\,.
\ee

   In an obvious notation, we may label the twelve positive-root 
generators by
\be
E_\a = (E_{\chi_1}, E_{\chi_2},\ldots, E_{\sigma_1}, E_{\sigma_2},\ldots,
    E_{\psi_1}, E_{\psi_2},\ldots)\,.\label{egens}
\ee
The simple root generators are $(E_{\chi_1},E_{\chi_2},E_{\chi_3},
E_{\sigma_4})$.  The non-vanishing commutators are given by
\bea
&&[E_{\sigma_1}, E_{\psi_1}]= E_{\chi_4}\,,\quad
[E_{\sigma_2}, E_{\psi_2}]= E_{\chi_4}\,,\quad
[E_{\sigma_3}, E_{\psi_3}]= E_{\chi_4}\,,\quad
[E_{\sigma_4}, E_{\psi_4}]= E_{\chi_4}\,,\\
&&[E_{\chi_2}, E_{\sigma_4}]= -E_{\sigma_1}\,,\quad
[E_{\chi_2}, E_{\sigma_3}]= E_{\sigma_2}\,,\quad
[E_{\chi_3}, E_{\sigma_1}]= -E_{\sigma_2}\,,\quad
[E_{\chi_3}, E_{\sigma_4}]= E_{\sigma_3}\,,\nn\\
&&[E_{\chi_3}, E_{\psi_2}]= E_{\psi_1}\,,\quad
[E_{\chi_1}, E_{\sigma_3}]= -E_{\psi_1}\,,\quad
[E_{\chi_1}, E_{\sigma_4}]= -E_{\psi_2}\,,\quad
[E_{\chi_2}, E_{\psi_2}]= -E_{\psi_3}\,,\nn\\
&&[E_{\chi_1}, E_{\sigma_1}]= -E_{\psi_3}\,,\quad
[E_{\chi_2}, E_{\psi_1}]= E_{\psi_4}\,,\quad
[E_{\chi_3}, E_{\psi_3}]= -E_{\psi_4}\,,\quad
[E_{\chi_1}, E_{\sigma_2}]= -E_{\psi_4}\,.\nn
\eea

   Following \cite{cjlp1} we now define a Borel-gauge 
coset representative ${\cal V}$ as follows:
\be
{\cal V} = e^{\ft12 \vec\varphi\cdot\vec H}\, {\cal U}_\chi\, {\cal U}_\sigma
\, {\cal U}_\psi\,,\label{vdef}
\ee
where
\bea
{\cal U}_\chi &=& e^{\chi_1\, E_{\chi_1}}\, e^{\chi_2\, E_{\chi_2}}\, 
                 e^{\chi_3\, E_{\chi_3}}\, e^{\chi_4\, E_{\chi_4}}\,,\nn\\
{\cal U}_\sigma &=& e^{\sigma_1\, E_{\sigma_1}}\, e^{\sigma_2\, 
  E_{\sigma_2}}\, e^{\sigma_3\, E_{\sigma_3}}\, 
                   e^{\sigma_4\, E_{\sigma_4}}\,,\nn\\
{\cal U}_\psi &=& e^{\psi_1\, E_{\psi_1}}\, e^{\psi_2\, 
  E_{\psi_2}}\, e^{\psi_3\, E_{\psi_3}}\, 
                   e^{\psi_4\, E_{\psi_4}}\,.
\eea
A straightforward calculation shows that if we define ${\cal U}\equiv
  {\cal U}_\chi\, {\cal U}_\sigma\, {\cal U}_\psi$ then 
\be
d{\cal U}\, {\cal U}^{-1} = {\cal F}= \sum_\a F^\a\, E_\a\,,
\ee
and we also have
\be
d{\cal V}\, {\cal V}^{-1} = \ft12 d\vec\varphi\cdot \vec H + 
       \sum_\a e^{\ft12 \vec a_\a\cdot\vec \varphi}\, F^\a\, E_\a\,.
\ee
Defining 
\be
{\cal M}\equiv {\cal V}^T\,\eta\,{\cal V}\,, \label{mdef}
\ee
it can be seen that the
fully-dualised three-dimensional Lagrangian (\ref{d3lag}) can be written 
as
\be
e^{-1}\, {\cal L}_3 = R - \ft18 \tr(\del{\cal M}^{-1}\, \del{\cal M})\,.
\ee
This makes the $O(4,4)$ global symmetry manifest.  Note that the
constant matrix $\eta$ in (\ref{mdef}) is chosen so that the required
distribution of positive and negative signs in the kinetic terms in
(\ref{d3lag}) is obtained.  Specifically, $\eta$ is preserved under an
$O(4,\C)$ subgroup of $O(4,4)$ matrices $K$:
\be
K^T\, \eta\, K = \eta\,.
\ee
(If we had instead performed a spacelike reduction to $D=3$, so that all 
the kinetic terms were of the standard sign, we would take $\eta=\oneone$,
and the subgroup of $O(4,4)$ matrices satisfying $K^T\, K=\oneone$ would
be $O(4)\times O(4)$.)

   In order to generate the 4-charge solution, we shall act on the 
dimensional reduction of the uncharged Kerr black hole with an 
$O(1,1)^4$ subgroup of $O(4,4)$.  Specifically, we shall take the 
$O(1,1)^4$ generators to be
\bea
&&\lambda_1= E_{\psi_1} + E_{\psi_1}^T\,,\qquad
\lambda_2= E_{\sigma_2} + E_{\sigma_2}^T\,,\nn\\
&&\lambda_3= E_{\psi_3} + E_{\psi_3}^T\,,\qquad
\lambda_4= E_{\sigma_4} + E_{\sigma_4}^T\,,\label{lambdadef}
\eea
and the $O(1,1)^4$ matrix
\be
\Lambda\equiv e^{\delta_i\, \lambda_i}
\ee
will be used to act on ${\cal V}$ by right multiplication.  In
principle, we can calculate the resulting transformations of the
fields from
\be
{\cal V}' = {\cal O}\, {\cal V}\, \Lambda\,,
\ee
where ${\cal O}$ is an $O(4,\C)$ compensating transformation that
restores the coset representative to the Borel gauge as in
(\ref{vdef}). In practice, the drawback to this approach is that
finding the required compensating transformation can be rather tricky.
Instead, we can calculate the field transformations using
\be
{\cal M}' = \Lambda^T\, {\cal M}\, \Lambda\,,\label{mtrans}
\ee
which avoids the need to find the compensating transformation.  The
price to be paid for this is that ${\cal M}$ is a much more
complicated matrix than ${\cal V}$.  However, by using an explicit
realisation for the $O(4,4)$ matrices (see appendix A), the problem
is easily tractable by computer.

\subsection{$O(1,1)^4$ transformation of a reduced uncharged solution}

   When we implement the solution-generating procedure, our starting
point in all cases will be an uncharged four-dimensional solution of
the ungauged ${\cal N}=2$ supergravity coupled to three vector
multiplets, whose bosonic equations of motion are described by the
Lagrangian (\ref{d4lag}).  More specifically, in all our examples the
starting point will be a solution of pure four-dimensional gravity,
\ie a Ricci-flat metric, admitting a timelike Killing vector
$\del/\del t$.  After reduction on the $t$ direction, it follows that
the only non-trivial three-dimensional fields will be the 3-metric
$ds_3^2$, the Kaluza-Klein vector $\cB_\1$ and the Kaluza-Klein scalar
$\varphi_4$.  The Kaluza-Klein vector is then dualised to an axion,
$\chi_4$, using (\ref{kkdual}).  In view of the fact that all the
scalars $\sigma_i$ and $\psi_i$ (associated with the reduction of the
4 four-dimensional vector fields) are zero in the starting
configuration, the dualisation of $\cB_\1$ at this stage is therefore
simply given by
\be
e^{2\varphi_4}\, {*d\cB_\1} = d\chi_4\,.\label{dual0}
\ee

   We now implement the $O(1,1)^4$ transformations, as described in 
section \ref{symsec}, taking as our starting point a three-dimensional 
configuration where only $ds_3^2$, $\varphi_4$ and $\chi_4$ are
non-trivial.  For convenience, we shall denote these starting expressions
for $\varphi_4$ and $\chi_4$ by $\td\varphi_4$ and $\td\chi_4$, and then
we denote the final expressions for all the $O(1,1)^4$-transformed fields
by their symbols without tildes.  (Since the three-dimensional metric
is inert under $O(4,4)$ transformations, we don't need to introduce
a tilde on the starting $ds_3^2$.)  

    The starting coset representative ${\cal V}$ is therefore given by
\be
{\cal V} = e^{\ft12 \td \varphi_4\, H_4}\, e^{\td\chi_4\, E_{\chi_4}}\,.
\ee
Constructing ${\cal M} = {\cal V}^T\, \eta\, {\cal V}$, and
acting with the $O(1,1)^4$ matrix $\Lambda$ as in (\ref{mtrans}), we
can obtain the transformed three-dimensional solution.  
Our results for the three-dimensional fields after $O(1,1)^4$
transformation are as follows:
\bea
\sigma_1 &=& \fft{\td\chi_4\Bigl( h_1 (c_{234} s_1 - c_1 s_{234}\,
e^{\td \varphi_4}) + c_1 s_1 s_{1234}\, \td\chi_4^2\Bigr)}{W^2}\,,\nn\\
\sigma_2 &=& \fft{c_2}{s_2} - \fft{c_2h_1 h_3 h_4 +
(c_{134}s_2 - c_2s_{134}\, e^{\td\varphi_4})\, s_{134}\,
\td\chi_4^2}{s_2W^2}\,,\nn\\
\sigma_3 &=& \fft{\td\chi_4\Bigl(h_3 (c_{124} s_3 - c_3 S_{124}\,
e^{\td \varphi_4}) + c_3s_3 s_{1234}\, \td\chi_4^2\Bigr)}{W^2}\,,\nn\\
\sigma_4 &=& \fft{c_4}{s_3} - \fft{c_4h_1 h_2 h_3 +
(c_{123}s_4 - c_4s_{123}\, e^{\td\varphi_4})\, s_{123}\,
\td\chi_4^2}{s_4W^2}\,,\nn\\
\psi_1 &=& \fft{c_1}{s_1} - \fft{c_1h_2 h_3 h_4 +
(c_{234}s_1 - c_1s_{234}\, e^{\td\varphi_4})\, s_{234}\,
\td\chi_4^2}{s_1W^2}\,,\nn\\
\psi_2 &=& -\fft{\td\chi_4\Bigl( h_2 (c_{134} s_2 - c_2 s_{134}\,
e^{\td \varphi_4}) + c_2 s_2 s_{1234}\, \td\chi_4^2\Bigr)}{W^2}\,,\nn\\
\psi_3 &=& \fft{c_3}{s_3} - \fft{c_3h_1 h_2 h_4 +
(c_{124}s_3 - c_3s_{124}\, e^{\td\varphi_4})\, s_{124}\,
\td\chi_4^2}{s_3W^2}\,,\nn\\
\psi_4 &=& -\fft{\td\chi_4\Bigl( h_4 (c_{123} s_4 - c_4 s_{123}\,
e^{\td \varphi_4}) + c_4 s_4 s_{1234}\, \td\chi_4^2\Bigr)}{W^2}\,,\nn\\
e^{\varphi_1} &=& \fft{h_1 h_3 + s_{13}^2\td\chi_4^2}{W}\,,\qquad
e^{\varphi_2}=\fft{h_2h_3 + s_{23}^2\td\chi_4^2}{W}\,,\nn\\
e^{\varphi_3}&=& \fft{h_1 h_2 + s_{12}^2 \td\chi_4^2}{W}\,,\qquad
e^{\varphi_4} = \fft{e^{\td\varphi_4}}{W}\,,\qquad
\chi_1 = \fft{(c_{13}s_{24} - c_{24}s_{13})\td\chi_4}{
h_1h_3 + s_{13}^2\td\chi_4^2}\,,\nn\\
\chi_2 &=& \fft{(c_{14}s_{23} - c_{23}s_{14})\td\chi_4}{
h_2h_3 + s_{23}^2\td\chi_4^2}\,,\qquad
\chi_3=\fft{(c_{12}s_{34} - c_{34}s_{12})\td\chi_4}{
h_1h_2 + s_{12}^2\td\chi_4^2}\,,\nn\\
\chi_4 &=& \fft{\td\chi_4}{W^2 (h_1h_2 + s_{12}^2\td\chi_4^2)}\Big\{
h_1 h_2\Bigl[ c_{1234} (1 + s_2^2 + s_4^2) +
s_{1234}(1 + s_2^2 + s_4^2)e^{2\td\varphi_4}\nn\\
&&- \Bigl( c_{1234} (s_2^2 +s_4^2) +
s_{1234} (2 + s_2^2 + s_4^2) \Bigr)e^{\td\varphi_4} \Bigr]
+s_{1234} s_{12}^2 (1 + s_2^2 + s_4^2) \td\chi_4^4\nn\\
&&+s_{12}\chi_4^2\Bigl[c_{12}(c_{12}s_{34} + c_{34}s_{12})(1 +
s_2^2 + s_4^2)-\Bigl( (c_{1234} s_{12} (s_2^2 + s_4^2)\nn\\
&& + s_{34} (s_1^2 +
s_2^2 + 5 s_{12}^2 + s_2^4 + 3 s_1^2 s_2^4 + s_1^2 s_4^2 +
s_{24}^2 + 3 s_{124}^2) \Bigr)e^{\td\varphi_4}\nn\\
&&+2s_{1234}s_{12} (1 + s_2^2 + s_4^2) e^{2\td\varphi_4}\Bigr]\Big\}\,,
\label{d3transformed}
\eea
where
\bea
h_i&=&c_i^2 -s_i^2\, e^{\td\varphi_4}\,,\quad
c_{i_1\cdots i_n}=\cosh\delta_i\,\cdots\,\cosh\delta_{i_n}\,,\quad
s_{i_1\cdots i_n}=\sinh\delta_i\,\cdots\,\sinh\delta_{i_n}\,,\nn\\
W^2&=&h_1h_2h_3h_4  + \td\chi_4^2 \Bigl(2 c_{1234}s_{1234} -
(s_{123}^2 + s_{124}^2 +s_{134}^2 + s_{234}^2 + 4 s_{1234}^2)
\, e^{\td\varphi_4}\nn\\
&&\qquad\qquad\qquad\quad  
  +2s_{1234}^2 e^{2\td\varphi_4}\Bigr) + s_{1234} \td\chi_4^4\,.
\eea

\section{4-Charge Rotating Black Holes in Ungauged Supergravity}

   In this section, we implement the procedure described in section
\ref{cgensec} to generate the solution for a 4-charge rotating black
hole in four dimensions.  Our starting point, therefore, is simply the
four-dimensional Kerr solution, namely
\bea
ds_4^2 &=&
  -dt^2 + \fft{2m\, r}{\rho^2}\, (dt-a\sin^2\theta d\varphi)^2 +
    (r^2+a^2)\sin^2\theta d\varphi^2 + \rho^2(\fft{dr^2}{\Delta} +
                                    d\theta^2)\,,\nn\\
\Delta &=& r^2 -2m r + a^2\,,\qquad \rho^2=r^2+a^2\cos^2\theta\,. 
\label{4dkerr}
\eea
Recasting it in the form (\ref{gred}), we can read off the reduced 
three-dimensional metric, Kaluza-Klein vector and Kaluza-Klein scalar:
\bea
ds_3^2 &=& (\rho^2-2mr)\Big(\fft{dr^2}{\Delta} + d\theta^2\Big) + 
   \Delta\, \sin^2\theta\, d\phi^2\,,\nn\\
\cB_\1 &=& \fft{2m \, a\, r\, \sin^2\theta\, d\phi}{\rho^2 -2m r}
\,,\qquad
   e^{\td\varphi_4} = 1 - \fft{2m r}{\rho^2}\,.
\eea
After dualisation, using (\ref{dual0}), the Kaluza-Klein 1-form
$\cB_\1$ becomes the axion $\td\chi_4$, which is given by
\be
\td\chi_4 = \fft{2m\, a\, \cos\theta}{\rho^2}\,.
\ee
All the other axions and dilatons in the three-dimensional theory
described by (\ref{d3lag}) are zero. Note that, in line with the
notation of section \ref{symsec}, we have placed tildes on the
starting expressions for the fields $\td\varphi_4$ and $\td\chi_4$.
The post-transformation fields are then written without tildes.

   After the $O(1,1)^4$ transformation, the fields are given by
(\ref{d3transformed}).  Before lifting the solution 
back to four dimensions, we must dualise the transformed
axions $\psi_i$ and $\chi_4$ back to 1-form potentials, so that we can
retrace the reduction steps.  After performing the dualisations in 
three dimensions, we find that the dilatons, axions and 1-form
potentials are given by 
\bea
\chi_1&=&\fft{2m\, u\, (c_{13}s_{24} -c_{24}s_{13})}{r_1\, r_3 + u^2}
\,,\qquad
\chi_2=\fft{2m\,u\,(c_{14}s_{23} - c_{23}s_{14})}{r_2\, r_3 + u^2}
\,,\nn\\
\chi_3&=&\fft{2m\,u\,(c_{12}s_{34} - c_{34}s_{12})}{r_1\, r_2 + u^2}
\,,\qquad
e^{\varphi_1} = \fft{r_1\, r_3 + u^2}{W}\,,\nn\\
e^{\varphi_2} &=& \fft{r_2\, r_3 + u^2}{W}\,,\quad
e^{\varphi_3} = \fft{r_1\, r_2 + u^2}{W}\,,\quad
e^{\varphi_4} = \fft{\rho^2 - 2m r}{W}\,,\nn\\
\sigma_1&=& \fft{2m\,u}{W^2}\Bigr[
(r \, r_1 + u^2) (c_{234}s_1-s_{234} c_1) +2m r_1 s_{234}c_1\Bigl]
\,,\nn\\
\sigma_2&=& \fft{1}{W^2}\Bigr[
2m\,c_2s_2\, (r_1\, r_3\, r_4 + r\, u^2) + 4 m^2\, u^2\, e_2 \Bigr]\nn\\
\sigma_3 &=& \fft{2m\,u}{W^2}\Bigl[(r\, r_3 + u^2) 
        (c_{124}s_3 - s_{124} c_3) + 2m r_3 s_{124}c_3\Bigr]\,,\nn\\
\sigma_4 &=& \fft{1}{W^2}\Bigl[2m\, c_4 s_4\, (r_1\, r_2\, r_3 + r\, u^2)
    + 4m^2\, u^2\, e_4 \Bigr]\,,\nn\\
\cB_\1 &=& \fft{2m(a^2-u^2)(rc_{1234}-
                      (r-2m)s_{1234})}{a(\rho^2 -2mr)}\,d\phi\,,\nn\\
A_{\1 1} &=& -\fft{2m\,u\, c_1s_1\,(r^2 + a^2 - 
2m r)}{a (\rho^2 - 2m r)}\,d\phi
\,,\nn\\
A_{\1 2} &=& \fft{2m(a^2-u^2)((r-2m)c_2s_{134}-
                      rc_{134}s_2)}{a(\rho^2 -2mr)}\,d\phi\,,\nn\\
\cA_\1^1 &=& -\fft{2m\,u\, c_3s_3\, 
(r^2 + a^2 - 2m\,r)}{a (\rho^2 - 2m r)}\,d\phi\,,\nn\\
\cA_\1^2 &=& \fft{2m(a^2-u^2)((r-2m)c_4s_{123}-
                      rc_{123}s_4)}{a(\rho^2 -2mr)}\,d\phi\,,\label{3dfields}
\eea
where
\bea
W^2&=&r_1\, r_2\, r_3\, r_4 + u^4 + u^2 [2r^2 + 2m r (s_1^2 + s_2^2
+ s_3^2 + s_4^2)\nn\\
&& + 8m^2 c_{1234}s_{1234} -
4m^2(s_{123}^2 + s_{124}^2 + s_{134}^2 +s_{234}^2 +2
s_{1234}^2)]\,,\nn\\
r_i &=& r + 2m s_i^2\,,\qquad u=a\cos\theta\,,\nn\\
c_{i_1\cdots i_n}&=& \cosh\delta_{i_1}\cdots\cosh\delta_{i_n}\,,\qquad
s_{i_1\cdots i_n} = \sinh\delta_{i_1}\cdots\sinh\delta_{i_n}\,.
\label{variousdefs}
\eea
We also define
\be
e_1 = c_{234}\, s_{234}\, (c_1^2 + s_1^2) - c_1\, s_1\, 
  (s_{23}^2 + s_{24}^2 + s_{34}^2 + 2 s_{234}^2)\,,\label{ddef}
\ee
together with analogous expressions for $e_2$, $e_3$ and $e_4$ defined
in the obvious way, with the label 2, 3 or 4 singled out in place
of the label 1.

\subsection{Four-dimensional 4-charge solution}

  The final step is to lift this three-dimensional solution back to
$D=4$, using the Kaluza-Klein reduction rules (\ref{gred}) and
(\ref{ared}). Thus the four-dimensional metric for the 4-charge
rotating black hole solution is given by
\be
ds_4^2 = -\fft{\rho^2-2mr}{W}\, (dt+ \cB_\1)^2 + 
    W\, \Big(\fft{dr^2}{\Delta} + d\theta^2 + 
   \fft{\Delta\, \sin^2\theta\, d\phi^2}{\rho^2-2mr}\Big)\,. \label{4dmetric}
\ee
The 4 four-dimensional gauge potentials, with we denote with hats here
to distinguish them from the three-dimensional ones, are given by
\bea
\hat A_{\1 1} &=& (A_{\1 1} + \sigma_1\, \cB_\1) + \sigma_1\, dt\,,\nn\\
\hat A_{\1 2} &=& (A_{\1 2} + \sigma_2\, \cB_\1) + \sigma_2\, dt\,,\nn\\
\hat \cA_{\1}^1 &=& (\cA_{\1}^1 + \sigma_3\, \cB_\1) + \sigma_3\, dt\,,\nn\\
\hat \cA_{\1}^2 &=& (\cA_{\1}^2 + \sigma_4\, \cB_\1) + \sigma_4\, dt\,.
\eea
The field strengths $\hat F_{\2 1}$ and $\hat \cF_\2^1$ carry magnetic
charges, while the field strengths $\hat F_{\2 2}$ and $\hat \cF_\2^2$ carry
electric charges.  The explicit expressions for the electric gauge
potentials are
\bea
\hat A_{1 2} &=& \fft{2m}{a W^2}\, \Big\{
  (r_1 \,r_3 \,r_4 + r \,u^2)[c_2 \,s_2 \,adt -(a^2-u^2)
(c_{134} \,s_2 - s_{134}\, c_2)\, d\phi] \nn\\
  && \qquad\qquad + 2m \,u^2 \,[e_2 \, adt - (a^2-u^2)\, s_{134} \,c_2
  d\phi]\Big\}\,,\nn\\
\hat \cA_{1}^2 &=& \fft{2m}{a W^2}\, \Big\{
  (r_1 \,r_2 \,r_3 + r \,u^2)[a \,c_4 \,s_4 \,dt 
   -(a^2-u^2)(c_{123} \,s_4 - s_{123}\, c_4)\,d\phi] \nn\\
  && \qquad\qquad + 2m \,u^2 \,[ a \,e_4 \,dt - 
    (a^2-u^2) s_{123}\, c_4 \,d\phi]\Big\}\,. \label{gaugep1}
\eea
The magnetic gauge potentials are more complicated; they take the form
\bea
\hat A_{\1 1} &=& \fft{2m\, u}{a W^2}\, 
\Big\{ (r\, r_1 + u^2)[(c_{234}\, s_1 - s_{234}\, c_1)\, a\, dt - c_1\,
s_1\, a^2\, d\phi] + 2m r_1\, s_{234}\, c_1\,a dt\nn\\
&& \qquad\qquad-\Big(c_1s_1(r_1r_2r_3r_4 + u^2[r^2 +2m\,r
(s_2^2 + s_3^2 + s_4^2)- 4m^2 s_{234}^2])\nn\\
&&\qquad\qquad\qquad +4m^2u^2 c_{234}s_{234}s_1^2
+ 2m e_1 (a^2 r_1 - r\,u^2)\Big)d\phi\Big\}
\,,\nn\\
\hat \cA_{\1 1} &=& \fft{2m\, u}{a W^2}\, 
\Big\{ (r\, r_3 + u^2)[(c_{124}\, s_3 - s_{124}\, c_3)\, a\, dt
- c_3\,s_3\, a^2\, d\phi] + 2m r_3\, s_{124}\, c_3\,a dt\nn\\
&& \qquad\qquad-\Big(c_3s_3(r_1r_2r_3r_4 + u^2[r^2 +2m\,r
(s_1^2 + s_2^2 + s_4^2)- 4m^2 s_{124}^2])\nn\\
&&\qquad\qquad\qquad +4m^2u^2 c_{124}s_{124} s_3^2 
+ 2m e_3 (a^2 r_3 - r\,u^2)\Big)d\phi\Big\}\,. \label{gaugep2}
\eea

     The remaining four-dimensional fields, namely the three dilatons
$(\varphi_1,\varphi_2,\varphi_3)$ and the three axions
$(\chi_1,\chi_2,\chi_3)$, are simply given by the three-dimensional
expressions in (\ref{3dfields}).  The results we have obtained here
complete those that were presented in \cite{CY4}, where the explicit
form of the four gauge potentials (\ref{gaugep1}) and (\ref{gaugep2})
was not given.

\subsection{Four-dimensional solution with pairwise-equal 
charges}\label{2ungaugedsec}

   A simple special case arises if we set the two electric charges
equal, by taking $\delta_4=\delta_2$, and also set the two magnetic
charges equal, by taking $\delta_3=\delta_1$.  We then find that the
previous expressions reduce as follows.  The function $W$ becomes
$W= r_1\, r_2 + a^2\, \cos^2\theta$,
and the four-dimensional metric is given by
\bea
ds_4^2 &=& -\fft{\rho^2-2mr}{r_1\, r_2 + a^2\, \cos^2\theta}\, 
\Big(dt - \fft{a\, \sin^2\theta\,  ( r^2- 2m r -r_1\, r_2)\, d\phi}{
\rho^2 - 2m r} \Big)^2 \nn\\
&& + (r_1\, r_2  + a^2\, \cos^2\theta)\, 
  \Big(\fft{dr^2}{\Delta} + d\theta^2 + 
   \fft{\Delta\, \sin^2\theta\, d\phi^2}{\rho^2-2mr}\Big)\,.
\eea
The remaining four-dimensional fields are given by
\bea
e^{\varphi_1} &=&\fft{r_1^2 + a^2\, \cos^2\theta}{r_1\, r_2 
                       + a^2\, \cos^2\theta}\,,\qquad
\chi_1^{\phantom{\chi_\chi}}=\fft{a\,(r_2-r_1)\, \cos\theta}{r_1^2 
                           + a^2\, \cos^2\theta}
= \fft{2m\, a\, (s_2^2-s_1^2)\, \cos\theta}{r_1^2+ a^2\, \cos^2\theta}  
\,,\nn\\
\hat A_{\1 1} &=& \hat\cA_\1^1= \fft{2m\, s_1\, c_1\, [a\, dt- 
  (r_1\, r_2 + a^2)\, d\phi]\, \cos\theta}{r_1\, r_2 + a^2\,
    \cos^2\theta}\,,\nn\\
\hat A_{\1 2} &=& \hat\cA_\1^2= \fft{2m\, s_2\, c_2\, r_1\, [dt-
   a\,\sin^2\theta\, d\phi]}{r_1\, r_2 + a^2\,
    \cos^2\theta}\,,\nn\\
\varphi_2&=&\varphi_3=\chi_2=\chi_3=0\,.\label{2chargesol}
\eea

   It can easily be verified that if one additionally sets
$\delta_1=\delta_2$, the solution reduces, as expected, to a dyonic
Kerr-Newman solution of the pure Einstein-Maxwell system, with equal
electric and magnetic charges.

    It is useful to write down the truncated four-dimensional theory for 
which the 2-charge configuration presented above is a solution.  The
truncation is performed by setting the field strengths in (\ref{d4lag})
pairwise equal (according to the scheme in (\ref{2chargesol}), and at
the same time setting $\varphi_2=\varphi_3=\chi_2=\chi_3=0$.  It is 
easily verified that this is a consistent truncation of the full equations
of motion.  The Lagrangian for the truncated system is
\bea
{\cal L}_4 &=& R\, {*\oneone} -\ft12 {*d\varphi_1}\wedge d\varphi_1 -
\ft12 e^{2\varphi_1}\, {*d\chi_1}\wedge d\chi_1 - \ft12 e^{-\varphi_1}\, 
 ({*F_{\2 1}}\wedge F_{\2 1} + {*F_{\2 2}}\wedge F_{\2 2}) \nn\\
&& - \ft12 \chi_1\, (F_{\2 1}\wedge F_{\2 1} + 
                           F_{\2 2}\wedge F_{\2 2})\,,\label{d4lag2}
\eea
where, having equated the pairs of field strengths, we have then 
rescaled them by factors of $1/\sqrt2$ in order to restore the 
canonical normalisation in (\ref{d4lag2}).  

   We can present the 2-charge solution in (\ref{2chargesol}) in a
more elegant form.  Recalling that we now have
\be
W = r_1\, r_2 + 
         a^2\, \cos^2\theta\,,
\ee
the terms in the metric can be reorganised so that it becomes 
\be
ds_4^2 = -\fft{\Delta}{W}\, (dt - a\, \sin^2\theta\,
d\phi)^2 + W\,  \Big( \fft{dr^2}{\Delta} + d\theta^2\Big) 
   + \fft{\sin^2\theta}{W}\, [a\, dt - (r_1 r_2 +
   a^2)d\phi]^2\,.\label{2charges1}
\ee
The remaining fields in the 2-charge solution may be written as
\bea
e^{\varphi_1} &=& 1 + \fft{r_1\, (r_1-r_2)}{W}= 
\fft{r_1^2 + a^2\, \cos^2\theta}{r_1\, r_2 
                       + a^2\, \cos^2\theta} \,,\qquad
   \chi_1 = \fft{a\, (r_2-r_1)\, \cos\theta}{r_1^2 + a^2\,
     \cos^2\theta}\,,\nn\\
A_{\1 1} &=& \fft{2\sqrt2\, m\, s_1\, c_1\, [a\, dt - (r_1 r_2 + a^2)
    d\phi]\, \cos\theta}{W}\,,\nn\\
A_{\1 2} &=& \fft{2\sqrt2 m\, s_2\, c_2\, r_1\, (dt-a\, \sin^2\theta\,
  d\phi)}{W}\,.\label{2charges2}
\eea

\section{Charged Rotating Black Holes in Gauged Supergravity}\label{gaugedsec}

   In this section, we look for generalisations of the charged rotating
black holes to the case of {\it gauged} four-dimensional supergravity.
Since there is no longer a solution-generating technique for deriving 
the solutions in the gauged theories, we instead resort to a technique of
``inspired guesswork,'' followed by brute-force verification that the 
equations of motion are satisfied.  The verification is purely mechanical, 
but the process of guessing, or making an ansatz, for the form of the
gauged solution is not so straightforward.  In fact, so far we have
succeeded in guessing the form of the gauged solution only in the
case that the four charges are set pairwise equal.  Thus in this
section, we shall present our results for the gauged generalisation
of the pairwise-equal ungauged solutions obtained in section
(\ref{2ungaugedsec}).

   The easiest way to discuss the solution is by simply augmenting the
bosonic Lagrangian (\ref{d4lag}) by the subtraction of the scalar
potential that arises in the gauged supergravity.  For the discussion 
of the solutions with pairwise-equal charges, which we are considering
here, we can take the scalar potential to be that of the ${\cal N}=4$ 
gauged $SO(4)$ theory, namely 
\be
  V= -g^2\, \sum_{i=1}^3\, (2 \cosh\varphi_i + \chi_i^2\,
  e^{\varphi_i})\,.\label{potdef}
\ee
with $\varphi_2=\varphi_3=\chi_2=\chi_3=0$.  The two electromagnetic 
charges in our pairwise-equal solution will then be carried by fields
in $U(1)$ subgroups of the two $SU(2)$ factors in $SO(4)\sim SU(2)\times 
SU(2)$.  Thus we may consider the bosonic Lagrangian
\bea
{\cal L}_4 &=& R\, {*\oneone} -\ft12 {*d\varphi_1}\wedge d\varphi_1 -
\ft12 e^{2\varphi_1}\, {*d\chi_1}\wedge d\chi_1 - \ft12 e^{-\varphi_1}\, 
 ({*F_{\2 1}}\wedge F_{\2 1} + {*F_{\2 2}}\wedge F_{\2 2}) \nn\\
&& - \ft12 \chi_1\, (F_{\2 1}\wedge F_{\2 1} + 
                           F_{\2 2}\wedge F_{\2 2})
-g^2\, (4 + 2 \cosh\varphi_1 + e^{\varphi_1}\, \chi_1^2)\, {*\oneone}
\,.\label{d4lag3}
\eea

    It should be emphasised that the Lagrangian (\ref{d4lag3}) is {\it not},
as it stands, the bosonic sector of any supergravity theory.  First of
all, we have included only the $U(1)\times U(1)$ subset of the $SU(2)\times 
SU(2)$ gauge fields of $SO(4)$-gauged ${\cal N}=4$ supergravity.  This
abelian truncation is consistent as a bosonic trunction, but not as a 
supersymmetric truncation.  Secondly, even if we included all the 
$SU(2)\times SU(2)$ gauge fields in (\ref{d4lag3}), the Lagrangian would
still not be the bosonic sector of the $SO(4)$-gauged ${\cal N}=4$ 
supergravity.  The gauge fields of one of the $SU(2)$ factors would
have had to have been dualised prior to turning on the gauging, in
order to get a supersymmetrisable theory.  Since bare potentials appear
in the expressions for the field strengths in the non-abelian gauged theory,
dualisation can no longer be performed.

   The upshot of the above discussion is that for the purposes of
conjecturing, and then verifying, a solution of the gauged theory, it
suffices to work with the generalisations of the ungauged solutions
(\ref{2chargesol}), and look at the equations of motion 
following from Lagrangian (\ref{d4lag3}).  Having successfully 
obtained charged rotating black-hole solutions, we can, if we wish, dualise
one of the two field strengths.  In this dualised form, the black hole
can be directly viewed as a solution within $SO(4)$-gauged ${\cal N}=4$
supergravity, with the non-zero gauge fields of the solution residing within
a $U(1)\times U(1)$ subgroup of $SU(2)\times SU(2)$.  It is this
dualised formulation that one would need to use if one wanted to test
the supersymmetry of the solution.

   By studying the form of the known Kerr-Newman-AdS black hole, 
as well that of the pairwise-equal charge solution (\ref{2chargesol}),
we have been able to conjecture the form of the rotating black hole
solution of gauged supergravity with two pair-wise 
equal charges.  Verifying the correctness of the conjecture is then 
a mechanical procedure, which we have performed using Mathematica. 
Our solution takes the form
\bea
ds_4^2 &=& -\fft{\Delta_r}{W}\, (dt - a\, \sin^2\theta\,
d\phi)^2 + W\,  \Big( \fft{dr^2}{\Delta_r} +
\fft{d\theta^2}{\Delta_\theta} \Big) 
   + \fft{\Delta_\theta\, \sin^2\theta}{W}\, [a\, dt - (r_1 r_2 +
   a^2)d\phi]^2\,,\nn\\
e^{\varphi_1} &=& =\fft{r_1^2 + a^2\cos^2\theta}{W}=
1 + \fft{r_1\, (r_1-r_2)}{W}\,,\qquad
   \chi_1 = \fft{a\, (r_2-r_1)\, \cos\theta}{r_1^2 + a^2\,
     \cos^2\theta}\,,\nn\\
A_{\1 1} &=& \fft{2\sqrt2\, m\, s_1\, c_1\, [a\, dt - (r_1 r_2 + a^2)
    d\phi]\, \cos\theta}{W}\,,\nn\\
A_{\1 2} &=& \fft{2\sqrt2 m\, s_2\, c_2\, r_1\, (dt-a\, \sin^2\theta\,
  d\phi)}{W}\,,\label{2charges4}
\eea
where $r_1$ and $r_2$ are defined in (\ref{variousdefs}), and
\bea
\Delta_r &\equiv& \Delta + g^2 \, r_1\, r_2\, (r_1\, r_2 + a^2)
=r^2 + a^2 - 2m\, r + g^2 \, r_1\, r_2\, (r_1\, r_2 +
a^2)\,,\nn\\
\Delta_\theta &\equiv& 1 - g^2\, a^2\, \cos^2\theta\,,\qquad
W=r_1\, r_2 + a^2 \cos^2\theta\,.
\eea
Note that the dilaton, axion and gauge potentials are identical to
those of the ungauged theory, appearing in (\ref{2charges2}).  

   As we mentioned above, it is useful for some purposes
to ``undualise'' the bosonic theory described by (\ref{d4lag3}), so
that it is expressed in terms of the fields that arise in the complete
$SO(4)$-gauged ${\cal N}=4$ supergravity including the fermions.  We can
do this by introducing a Lagrange multiplier $B_\1$ to enforce the
Bianchi identity $dF_{\2 1}=0$, adding the term $dB_\1\wedge F_{\2 1}$
to the Lagrangian (\ref{d4lag3}), and using the algebraic equation of
motion for $F_{\2 1}$ to eliminate $F_{\2 1}$ in favour of $G_\2=dB_\1$.
Thus we find
\be
e^{-\varphi_1}\, {*F_{\2 1}} + \chi_1\, F_{\2 1} = G_\2\,,
\ee
and hence the Lagrangian becomes
\bea
{\cal L}_4 &=& R\, {*\oneone} -\ft12 {*d\varphi_1}\wedge d\varphi_1 -
\ft12 e^{2\varphi_1}\, {*d\chi_1}\wedge d\chi_1 - \ft12 e^{-\varphi_1}\, 
 {*F_{\2 2}}\wedge F_{\2 2} -\ft12 \chi_1\, F_{\2 2}\wedge F_{\2 2} \nn\\
&& -\fft1{2(1+\chi_1^2\, e^{2\varphi_1})}\, (
   e^{\varphi_1}\, {*G_\2}\wedge G_\2 - e^{2\varphi_1}\, \chi_1\, 
       G_\2\wedge G_\2)\nn\\
&&
-g^2\, (4 + 2 \cosh\varphi_1 + e^{\varphi_1}\, \chi_1^2)\, {*\oneone}
\,.\label{d4lag4}
\eea

   In terms of this ``undualised'' formulation, the 2-charge solution
(\ref{2charges4}) is identical except that instead of having $A_{\1 1}$ 
which carries magnetic charge, we have 
\be
B_\1 = \fft{2\sqrt2 m\, s_1\, c_1\, r_2\, (dt-a\, \sin^2\theta\,
  d\phi)}{W}\,,\label{bfield}
\ee
which carries electric charge.

     It is easy to see that when $\delta_1=\delta_2$, the dilaton
$\phi_1$ and the axion $\chi_1$ in (\ref{2charges4}) vanish, and the
solution reduces to the AdS-Kerr-Newman black hole with purely electric
charge \cite{carter1,carter2}.  If instead we set the rotation parameter $a$
to zero, the solution becomes the static AdS black hole constructed
in \cite{duffliu}, with the four charges set pairwise equal.

    It was shown in \cite{s7red} that the ${\cal N}=4$ gauged
supergravity (\ref{d4lag3}) can be obtained from $S^7$ reduction
of eleven-dimensional supergravity, where explicit reduction ansatz
were given.  We can use the reduction ansatz to lift the charged
rotating solution back to $D=11$.  The metric becomes
\cramper{
\bea
ds_{11}^2 \!\!\!&=&\!\!\! (X^2\cos^2\xi + \sin^2\xi)^{\ft13}
(\wtd X^2\sin^2\xi + \cos^2\xi)^{\ft13}
\Bigl\{-\fft{\Delta_r}{W}\, (dt - a\, \sin^2\theta\,
d\phi)^2\nn\\
\!\!\!&&\!\!\!
+ W\,\Big( \fft{dr^2}{\Delta_r} +\fft{d\theta^2}{\Delta_\theta} \Big)
+ \fft{\Delta_\theta\, \sin^2\theta}{W}\, 
[a\, dt - (r_1 r_2 + a^2)d\phi]^2 + 4g^{-2}\, d\xi^2\\
\!\!\!&&\!\!\! +\fft{\cos^2\xi [
d\theta_1^2 + \sin^2\theta_1\, d\phi_1^2 + (d\psi_1 +
\cos\theta_1\, d\phi_1 - 2 m\,g\, s_2\, c_2\, r_1 W^{-1}(dt - a
\sin^2\theta\, d\phi)^2]}{g^2(X^2\cos^2\xi + \sin^2\xi)}
\nn\\
\!\!\!&&\!\!\! +\fft{\sin^2\xi [
d\theta_2^2 + \sin^2\theta_2\, d\phi_2^2 + (d\psi_2 +
\cos\theta_2\, d\phi_2 - 2 m\,g\, s_1\, c_1\, r_2 W^{-1}(dt - a
\sin^2\theta\, d\phi)^2]}{g^2(\wtd X^2\sin^2\xi + \cos^2\xi)}\Bigr\}
\,,
\nn
\eea
}
where 
\be
X^2=\fft{r_1^2 + a^2\cos^2\theta}{W}\,,\qquad
\wtd X^2 = \fft{r_2^2 + a^2\cos^2\theta}{W}
\,,
\ee
and $(\theta_1, \phi_1, \psi_1)$ and $(\theta_2, \phi_2, \psi_2)$ are
Euler angles on the two $S^3$ factors of the foliation of $S^7$.  The
eleven-dimensional metric describes a rotating M2-brane with rotations
in both the world-volume and the transverse space.  The 4-form field
strength can be easily read off from the reduction formulae in
\cite{s7red}.  If we set the world-volume rotation parameter $a$ to
zero, the eleven-dimensional solution reduces to the one obtained in
\cite{tenauthor}, where the four transverse angular momenta are set
pairwise equal.

\section{Generalisations with a NUT Parameter}\label{nutsec}

   The construction of four-dimensional charged solutions that we have
been discussing so far were obtained by starting from the uncharged
Kerr solution, and performing an $O(1,1)^4 \subset O(4,4)$ transformation
in a dimensional reduction to three dimensions.  Lifting back to four
dimensions, this yields a solution of ungauged supergravity with four
independent charges, supported by four abelian field strengths.  In
the case where the four charges were set pairwise equal, we were able
to conjecture the generalisation of this solution to gauged
supergravity, and to confirm that indeed the equations of motion were
satisfied.

   More generally, we can start from an uncharged solution where not
only the rotation parameter, but also the NUT charge parameter, is
non-zero.  Reducing this Ricci-flat Kerr-Taub-NUT solution to $D=3$,
performing the $O(1,1)^4$ transformation, and lifting back to $D=4$
yields a new solution of ungauged supergravity, with four
electromagnetic charges, rotation, mass and NUT charge.  Again, in the
case that the four electromagnetic charges are set pairwise equal, we
can successfully conjecture a generalisation of the solution to gauged
supergravity, and verify that indeed it solves the equations.  

   The procedure for first constructing the generalisations of the
Kerr-Taub-NUT metrics to include 4 charges is analogous
to the one we followed previously for the Kerr metric.  A convenient
way of writing the Kerr-Taub-NUT metric, which forms the starting point
for our procedure, is in the formulation of Plebanski \cite{pleb}:
\be
ds_4^2 = - \fft{\wtd\Delta_r}{a^2(r^2+u^2)}\,  [a dt
+ u^2\, d\phi]^2 +\fft{\wtd\Delta_u}{a^2(r^2+u^2)}\,  
[adt - r^2\, d\phi]^2 + 
 (r^2+u^2)\,\Big(\fft{dr^2}{\wtd\Delta_r} +
\fft{du^2}{\wtd\Delta_u}\Big)\,,\label{plebmet}
\ee
where the functions $\wtd\Delta_r$ and $\wtd\Delta_u$ are given by
\be
\wtd\Delta_r = a^2 +r^2 -2m r\,,\qquad
\wtd\Delta_u = a^2 -u^2 + 2\ell\, r\,.
\ee
Here $m$ is the mass, $\ell$ is the NUT parameter, and $a$ is the
rotation parameter.

   We then implement the reduction (on the time coordinate $t$) to
three dimensions, and performing the $O(1,1)^4$ rotation as described
in section (\ref{cgensec}).  After ``undualising'' scalars to vectors,
we lift back to four dimensions, thereby obtaining a solution with
four electromagnetic charges.  As in our earlier discussion for the
Kerr case, two charges, associated with the $O(1,1)^4$ parameters
$\delta_1$ and $\delta_3$, are magnetic, while the two associated with
$\delta_2$ and $\delta_4$ are electric.  

     In the present section, we give our results for the case
where we set the charges pairwise equal, \ie $\delta_1=\delta_3$ and
$\delta_2=\delta_4$.  Since the ``inspired guesswork'' that leads to the
generalisation of the solution to the gauged theory is very similar
to that which we used in section (\ref{gaugedsec}) for the case with no NUT
parameter, we shall in fact directly present our results for the gauged case.
We find that the metric of the charged
four-dimensional solution is given by
\be
ds_4^2 = - \fft{\Delta_r}{a^2\, W}\,  [a dt
+ u_1\, u_2\, d\phi]^2 +\fft{\Delta_u}{a^2\, W}\,  
[adt - r_1\, r_2\, d\phi]^2 + 
  W\,\Big(\fft{dr^2}{\Delta_r} +
\fft{du^2}{\Delta_u}\Big)\,,
\ee
where 
\bea
\Delta_r &=& \wtd\Delta_r + g^2\, r_1\, r_2\, (r_1\, r_2 +
a^2)\,,\qquad
\Delta_u = \wtd\Delta_u + g^2\, u_1\, u_2\, (u_1\, u_2 -
a^2)\,,\nn\\
W&=& r_1\, r_2 + u_1\, u_2\,,\qquad
r_i= r + 2m\, s_i^2\,,\qquad u_i = u + 2\ell\, s_i^2\,,
\eea
and $s_i=\sinh \delta_i$.  The remaining fields are given by
\bea
e^{\varphi_1} &=& \fft{r_1^2 + u_1^2}{W}\,,\qquad 
 \chi_1 = \fft{2(s_1^2-s_2^2)\, (\ell\, r -m\,
   u)}{r_1^2+u_1^2}\,,\nn\\
A_{\1 1} &=& \fft{2 \sqrt2 s_1\, c_1}{a\, W}\, \Big\{
    m\, u_1\, [a dt - r_1\, r_2\,  d\phi] - \ell\, r_1\, 
 [a dt + u_1\, u_2\, d\phi]\Big\}\,,\nn\\
A_{\1 2} &=& \fft{2\sqrt2  s_2\, c_2}{a\, W}\, \Big\{
    \ell\, u_1\, [a dt - r_1\, r_2\,  d\phi] +m \, r_1\, 
 [a dt + u_1\, u_2\, d\phi]\Big\}\,.
\eea
The ungauged case is, of course, 
obtained by setting $g=0$.  The verification of our conjectured 
result for $g\ne0$ is straightforward; we used Mathematica to check that the
equations of motion following from (\ref{d4lag4}) are indeed satisfied.
In Appendix B, we give our complete results for the rotating NUT solutions
with four independent charges, in the ungauged case.

   It can easily be seen that this solution reduces to the previous
one given in (\ref{2charges4}) if one sets the NUT parameter $\ell$ to
zero.  If instead the two charge parameters are set equal,
$\delta_1=\delta_2$, then one has $\varphi_1=\chi_1=0$ and the
solution reduces to the charged AdS-Kerr-Taub-NUT solution of
Einstein-Maxwell theory with a cosmological constant, as given in
\cite{pleb}.

       As in the case of the rotating black hole in the previous chapter,
we can lift the solution back to $D=11$, following the reduction ansatz
given in \cite{s7red}.

\section{Generalisation with Acceleration}

       In addition to the inclusion of a NUT parameter, there exist 
more general four-dimensional type-D metrics which include an acceleration
parameter as well as mass, NUT charge and rotation.  These metrics can
be elegantly described within the framework of Plebanski and Demianski
\cite{plebdem}, where they are written as
\be
ds^2 =\fft{1}{(1 - r u)^2} \Big\{\fft{\Delta_u}{r^2 + u^2}
(dt - r^2 d\phi)^2 - \fft{\Delta_r}{r^2 + u^2} (dt + u^2 d\phi)^2
+ (r^2 + u^2)(\fft{dr^2}{\Delta_r} + \fft{du^2}{\Delta_u})\Big\}\,.
\ee
Here 
\bea
\Delta_r &=& (\gamma -\ft16\Lambda) - 2m\, r + \epsilon\, r^2 -
2\ell\, r^3 - (\gamma  + \ft16\Lambda ) r^4\,,\nn\\
\Delta_u &=& (\gamma  -\ft16\Lambda ) + 2\ell\, u - \epsilon\, u^2 +
2m\, u^3 - (\gamma  + \ft16\Lambda) u^4\,.
\eea
We have included here the cosmological constant $\Lambda$; in order to
implement the charge-generating procedure of section (\ref{cgensec}),
we shall, of course, need to set $\Lambda=0$.  In fact the general
Plebanski-Demianski metrics also include electric and magnetic
charges, giving solutions of the Einstein-Maxwell equations. Our
interest is in constructing more general solutions in four-dimensional
supergravities, along the lines we have considered in previous
sections.  Thus for our purposes we can begin with the uncharged
Plebanski-Demianski metrics. We also turn off the cosmological
constant $\Lambda$, so that we can apply the solution-generating
technique via $O(1,1)^4$ transformations after reduction to three
spatial dimensions.

   We then perform the $O(1,1)^4$ boosts in three dimensions, undualise 
the scalars that originally corresponded to vectors, and lift
the solution back to $D=4$, where we can obtain 4-charged rotating and
accelerating solutions with mass and NUT charge.  We have worked out
the solutions with four independent charges, but they are rather
involved and so, for simplicity, we
shall present here only the solution where charges are set pairwise equal,
namely $\delta_3=\delta_1$ and $\delta_4=\delta_2$.  We find that the
solution is given by
\bea
ds^2 &=&\fft{1}{(1 - r u)^2} \Big\{\fft{\Delta_u}{W}
(dt - r_1 r_2 d\phi)^2 - \fft{\Delta_r}{W} (dt + u_1u_2 d\phi)^2
+ W(\fft{dr^2}{\Delta_r} + \fft{du^2}{\Delta_u})\Big\}\,,\nn\\
e^{\phi_1} &=& \fft{r_1^2 + u_1^2}{W}\,,\qquad
\chi_1=\fft{(s_1^2 - s_2^2) (r_1 u_s - u_1 r_s)}{r_1^2 + u_1^2}
\,,\nn\\
A_\1^1 &=& \fft{s_1 c_1}{(1-r u)^2\, W}\Big\{
f_1 (dt - r_1 r_2 d\phi) + f_2 (dt + u_1 u_2 d\phi) \Big\}\,,\nn\\
A_\1^2 &=& \fft{s_2 c_2}{(1 - r u)^2 W}\Big\{ u_1 r_s (
dt - r_1 r_2 d\phi) + r_1 r_s (dt + u_1 u_2 d\phi)\Big\}\,,
\eea
where
\bea
r_i &=& r + r_s\, s_i^2\,,\qquad u_i = u + u_s\, s_i^2\,,
\qquad W=r_1 r_2 + u_1 u_2\,,\nn\\
r_s &=& \fft{1}{r} (r^2 + \gamma - \fft{\Delta_r}{(1-r u)^2})
\,,\qquad u_s = \fft{1}{u} (u^2 - \gamma +
\fft{\Delta_u}{(1- r u)^2})\,,\nn\\
\Delta_r&=&\gamma - 2m\,r + \epsilon\, r^2 - 2\ell\, r^3 -
\gamma\, r^4\,,\nn\\
\Delta_u&=&\gamma + 2\ell\, u - \epsilon\, u^2 + 2m u^3 -
\gamma\, u^4\,,\nn\\
f_1 &=& \gamma\, r^2 - 2(m-\ell\, r^2) u +
(\gamma + 4m\, r - \epsilon\, r^2) u^2 - 2\gamma\, r\, u^3\,,\nn\\
f_2 &=& \gamma\, u^2 + 2(\ell - m\, u^2) r +
(\gamma - 4\ell\, u + \epsilon\, u^2) r^2 - 2\gamma\, u\, r^3\,.
\eea

  In our previous examples, of charged generalisations of the Kerr and 
Kerr-Taub-NUT metrics, we found that the necessary modifications in order
to obtain solutions within {\it gauged} supergravity required 
changing only the metric, simply by
adding terms to the functions $\Delta_r$ and $\Delta_u$. 
The dilaton, the axion and the two gauge potentials remained unchanged.
In the present case, however, these fields would also have to be modified.
To see this, we can examine the equations of motion for the
vector potentials.  

   We could only expect that the gauge potentials would require no
modification when seeking solutions in the gauged supergravity if the
gauge-field equations of motion did not involve the explicit
appearance of the functions $\Delta_r$ and $\Delta_u$ that themselves
are modified in the gauged case.  In order for the $\Delta_r$ and
$\Delta_u$ functions to be absent in the gauge-field equations, it is 
necessary that the field strengths take the form
\be
F_\2 = \alpha_1 dr\wedge (dt + u_1 u_2 d\phi) + \alpha_2
du\wedge (dt - r_1 r_2 d\phi)\,,
\ee
where $\alpha_1$ and $\alpha_2$ are functions of $r$ and $u$.
For gauge potential of the form $A_\1 = \beta_1 dt + \beta_2 d\phi$,
we must then have the conditions
\be
\fft{\del \beta_2}{\del r} - u_1u_2 \fft{\del\beta_1}{\del r}=0\,,\qquad
\fft{\del \beta_2}{\del u} + r_1r_2 \fft{\del\beta_1}{\del u}=0\,.
\label{betacon}
\ee
It is straightforward to verify that the above conditions are
satisfied by the previous examples (charged generalisations of Kerr or
Kerr-Taub-NUT), and consequently, the vector potentials receive no
modification when the solutions are generalised to gauged
supergravity, which is achieved by adding terms only to $\Delta_r$ and
$\Delta_u$.  In the present example, however, the conditions
(\ref{betacon}) are not satisfied, and so we expect that the
generalisation to gauged supergravity would require modifications
to the gauge potentials and, possibly, to the dilaton and axion, 
as well as to the metric.

\section{Conclusions}

    In this paper we have constructed new charged rotating solutions
of four-dimensional ungauged and gauged supergravities.  
Our new ungauged solutions can be viewed as being embedded 
within ${\cal N}=2$ supergravity coupled to three vector multiplets.
This is itself, of course, embedded within ${\cal N}=8$ supergravity.
We first constructed the general ungauged rotating black holes with
four independent charges; this completed the presentation of the
solutions first given in \cite{CY4}, which did not give the expressions
for the gauge potentials.   We did this by employing a solution-generating 
technique, which involved reducing the four-dimensional theory on 
the time direction and then acting with global symmetry generators
$O(1,1)^4\subset O(4,4)$ to introduce the charges.  We were able also
to construct four-dimensional charged generalisations of the 
Kerr-Taub-NUT solution, and the further generalisation where the 
acceleration parameter is included too.  In all cases we constructed 
the solutions with four independent charge parameters, although in the
case of the rotating metrics with acceleration parameter as well as
the NUT parameter, the general 4-charge solutions were rather too
complicated to include in the paper.  In that case, we therefore only 
presented the specialisation where the four charges are set pairwise
equal.

   For the case of the charged rotating black holes, both with 
and without the inclusion of the NUT parameter, we were able to conjecture
the generalisations of the above ungauged solutions to the case of
gauged supergravity, after making the specialisation that the four charges
are set pairwise equal.  We verified the correctness of our conjectured
solutions by explicitly confirming that all the equations of motion
are satisfied.  These solutions are most appropriately
viewed as being embedded in $SO(4)$-gauged ${\cal N}=4$ supergravity.

   The four-dimensional charged rotating black hole solutions that we
obtained in this paper provide new gravitational backgrounds for
four-dimensional vacua in compactified string theory.  In particular,
the non-extreme black hole solutions of gauged supergravity provide
asymptotically AdS backgrounds that are characterised by their mass,
angular momentum and two pair-wise equal charges (implying that they
can be viewed as solutions in ${\cal N}=4$ gauged supergravity).  The
more general gauged solutions that we also obtained, where
additionally the NUT parameter is non-zero, should provide new
information on the dual three-dimensional conformal field theory at
non-zero temperature.

\section*{Acknowledgments}

M.C. and C.N.P.  would like to thank CERN for 
hospitality and support during the initial stages
 of the work.

\appendix

\section{Matrix Realisation of the $O(4,4)$ Generators}

   The $O(4,4)$ algebra has an $SL(4,\R)$ subalgebra, whose
positive-root generators are $E_i{}^j$, satisfying
\be
[E_i{}^j, E_k{}^\ell] = \delta_k^j\, E_i{}^\ell - \delta_i^\ell\, 
      E_k{}^j\,.\label{ealg}
\ee
To this we append the generators $V^{ij}$, antisymmetric in $i$ and
$j$, which satisfy the commutation relations
\be
[E_i{}^j, V^{k\ell}] = -\delta_i^k\, V^{j\ell} - \delta_i^\ell\,
V^{kj}\,,\qquad [V^{ij}, V^{k\ell}]=0\,.\label{valg}
\ee
Together, $(E_i{}^j,V^{ij})$ form the positive-root generators of 
$O(4,4)$.  The simple-root generators are
\be
V^{12}\,,\qquad E_1{}^2, \qquad E_2{}^3\,,\qquad E_3{}^4\,.
\ee
In terms of our generators in (\ref{egens}), we have
\be
E_{\chi_1} =V^{12}\,,\qquad 
E_{\chi_2} = E_1{}^2\,,\qquad
E_{\chi_3} =E_3{}^4\,,\qquad
E_{\sigma_4} = E_2{}^3\,.
\ee
The remaining positive-root generators in section xxx are given by
\bea
&&E_{\sigma_1}= -E_1{}^3\,,\qquad E_{\sigma_2}= -E_1{}^4\,,\qquad
E_{\sigma_3} = -E_2{}^4\,, \qquad E_{\chi_4} = V^{34}\,,\nn\\
&& E_{\psi_1}= V^{14}\,,\qquad
  E_{\psi_2} = -V^{13}\,,\qquad
   E_{\psi_3}= -V^{23}\,,\qquad
    E_{\psi_4}= - V^{24}\,.
\eea

   To give an explicit $8\times 8$ matrix realisation, we first introduce
the $4\times 4$ matrix $e_{ij}$ which has zeros everywhere except for
a 1 at row $i$, column $j$.  We can then write
\be
E_i{}^j = \left(\begin{array}{c|c}
                  -e_{ji} & 0 \\ \hline
                     0 & e_{ij}
                    \end{array}\right)\,,\qquad
V^{ij} = \left(\begin{array}{c|c}
                   0 & e_{ij}-e_{ji} \\ \hline
                     0 & 0
                    \end{array}\right)\,,
\ee
where each block represents a $4\times 4$ matrix.  The four Cartan
generators are then represented by
\bea
&&H_1 = \hbox{diag}(1,1,0,0,-1,-1,0,0)\,,\qquad
H_2 = \hbox{diag}(-1,1,0,0,1,-1,0,0)\,,\nn\\
&&H_3 = \hbox{diag}(0,0,-1,1,0,0,1,-1)\,,\qquad
H_4 = \hbox{diag}(0,0,-1,-1,0,0,1,1)\,.
\eea
The negative-root generators are given by the matrix transposes of the
positive-root generators.

All of the above discussion generalises straightforwardly to $O(n,n)$,
with its $SL(n,\R)$ subgroup having positive-root generators $E_i{}^j$
($i<j$), and the full $O(n,n)$ obtained by augmenting these with
$V^{ij}$.  The algebra of the $E_i{}^j$ and $V^{ij}$ is given by
(\ref{ealg}) and (\ref{valg}), with $1\le i\le n$.  The simple-root
generators of $O(n,n)$ are the simple-root generators $E_i{}^{i+1}$ of
$SL(n,\R)$, augmented by $V^{12}$.

   The matrix $\eta$, used in defining ${\cal M}$ in
(\ref{mdef}), as a coset representative for $O(4,4)/O(4,\C)$, is given
in this basis by
\be
\eta=\hbox{diag}(1,1,-1,-1,1,1,-1,-1)\,.
\ee
The $O(4,\C)$ subgroup comprises $O(4,4)$ matrices $K$ that satisfy 
$K^T\, \eta\, K=\eta$, and so the $O(4,\C)$ generators $T_\a$ 
comprise the subset of $O(4,4)$ generators that satisfy $T_\a^T\, \eta + 
\eta\, T_\a =0$.  It is straightforward to see that these are
\be
X_i=E_{\sigma_i} + E_{\sigma_i}^T\,,\qquad
Y_i=E_{\psi_i} + E_{\psi_i}^T\,,\qquad
Z_i=E_{\chi_i} - E_{\chi_i}^T\,,\label{xyzdef}
\ee
where $1\le i\le 4$. The generators $X_i$ and $Y_i$ are non-compact,
whilst the generators $Z_i$ are compact.  

    It is convenient to take the Cartan generators $h_i$ to be combinations
of the four
generators $(\lambda_1, \lambda_2,\lambda_3,\lambda_4)$ that were
defined in (\ref{lambdadef}), which we exponentiated with parameters
$\delta_i$ to give the $O(1,1)^4$ charge-generating transformations.
In the notation of (\ref{xyzdef}), we have
\be
\lambda_1 = Y_1\,,\qquad \lambda_2=X_2\,,\qquad \lambda_3=Y_3\,,\qquad
   \lambda_4=X_4\,.\label{lambdas}
\ee
We define 
\bea
&&h_1=\ft12(\lambda_1+\lambda_2+\lambda_3+\lambda_4)\,,\qquad
h_2=\ft12(\lambda_1-\lambda_2-\lambda_3+\lambda_4)\,,\nn\\
&&h_3=\ft12(\lambda_1-\lambda_2+\lambda_3-\lambda_4)\,,\qquad
h_4=\ft12(\lambda_1+\lambda_2-\lambda_3-\lambda_4)\,.
\eea
A straightforward calculation shows that the positive and negative
root generators of $O(4,\C)$ are then given by
\be
E^\pm_i = U_i \pm V_i\,,
\ee
where 
\bea
&& U_1 = X_1 + X_3 -Y_2 -Y_4\,,\qquad
   V_1 = -Z_1 + Z_2 - Z_3 - Z_4\,,\nn\\
&& U_2 = X_1 - X_3 +Y_2 -Y_4\,,\qquad
   V_2 = -Z_1 - Z_2 - Z_3 + Z_4\,,\nn\\
&& U_3 = X_1 - X_3 -Y_2 +Y_4\,,\qquad
   V_3 = Z_1 - Z_2 - Z_3 - Z_4\,,\nn\\
&& U_4 = X_1 + X_3 +Y_2 +Y_4\,,\qquad
   V_4 = Z_1 + Z_2 - Z_3 + Z_4\,.
\eea
Under $\vec h\equiv (h_1,h_2,h_3,h_4)$, 
the roots have weights given by $[\vec h, E^\pm_i] = \pm \vec
\a_i\, E^\pm_i$, where 
\be
\vec\a_1=(2,0,0,0)\,,\quad \vec\a_2=(0,2,0,0)\,,\quad
\vec\a_3=(0,0,2,0)\,,\quad \vec\a_4=(0,0,0,2)\,.
\ee
In fact the $O(4,\C)$ algebra can now be seen to be nothing but the
sum of four mutually-commuting copies of the $SL(2,\R)$ algebra,
generated by $(E^+_i, E^-_i, h_i)$ for each $1\le i\le 4$.

\section{Four-charge rotating NUT solution}

   In this Appendix, we present the general 4-charge solution obtained
by applying the $O(1,1)^4$ transformation to the Plebanski metric
(\ref{plebmet}).  This generalises the pairwise-equal case presented
in section \ref{nutsec}.

    Making the definitions
\bea
r_0\,=\,r-2\,m\,,\qquad  u_0\,=\,r-2\,\ell\,,\qquad
\bar\rho\,=\,r\,r_0\,+\,u\,u_0\,,
\eea
and
\bea
W^2\,&=&\, r_1\,r_2\,r_3\,r_4\,+\,u_1\,u_2\,u_3\,u_4\,+
\,2\,u^2\,r^2\,+\,2\,r\,u\,(\ell\,r\,+\,m\,u)(s_1^2\,+\,s_2^2\,+
\,s_3^2\,+\,s_4^2)\,\nn\\
&-&\,4\,(\ell\,r\,-\,m\,u)^2(s_{123}^2\,+\,s_{124}^2\,+\,s_{134}^2\,+
\,s_{234}^2\,+\,2\,s_{1234}^2\,-\,2\,c_{1234}\,s_{1234})\nn\\
&+&\,8\,m\,\ell\,r\,u\,(s_{12}^2\,+\,s_{13}^2\,+\,s_{14}^2\,+
\,s_{23}^2\,+\,s_{24}^2\,+\,s_{34}^2)\nn\\
&+&\,8\,m\,\ell(m\,u +\ell\, r)\,(s_{123}^2\,+\,s_{124}^2\,+
\,s_{134}^2\,+\,s_{234}^2)\,+\,32\,m^2\,\ell^2\,s_{1234}^2
\eea
we find that in three dimensions the 4-charge solution is given by
\bea
\c_1\,&=&\,\fft{2\,(c_{24}\,s_{13}-c_{13}\,s_{24})\,(\ell\,r\,-
\,m\,u)}{r_1\,r_3\,+\,u_1\,u_3},\quad
\c_2\,=\,\fft{2\,(c_{14}\,s_{23}-c_{23}\,s_{14})\,(\ell\,r\,-
\,m\,u)}{r_2\,r_3\,+\,u_2\,u_3},\nn\\
\c_3\,&=&\,\fft{2\,(c_{34}\,s_{12}-c_{12}\,s_{34})\,(\ell\,r\,-
\,m\,u)}{r_1\,r_2\,+\,u_1\,u_2},\quad
e^{\varphi_1}\,=\,\fft{r_1\,r_3\,+\,u_1\,u_3}{W},\nn\\
e^{\varphi_2}\,&=&\,\fft{r_2\,r_3\,+\,u_2\,u_3}{W},\quad
e^{\varphi_3}\,=\,\fft{r_1\,r_2\,+\,u_1\,u_2}{W},\quad
e^{\varphi_4}\,=\,\fft{r\,r_0\,+\,u\,u_0}{W},\nn\\
\s_1\,&=&\,\fft{2}{W^2}\,(\ell\,r\,-\,m\,u)\,[c_1\,s_{234}\,
(r_0\,r_1\,+u_0\,u_1)\,-\,s_1\,c_{234}\,(r\,r_1\,+\,u\,u_1)],\nn\\
\s_3\,&=&\,\fft{2}{W^2}\,(\ell\,r\,-\,m\,u)\,[c_3\,s_{124}\,
(r_0\,r_3\,+u_0\,u_3)\,-\,s_3\,c_{124}\,(r\,r_3\,+\,u\,u_3)],\nn\\
\s_2\,&=&\,\fft{2}{W^2}\,\Big [c_2\,s_2\,\Big(m\,r_1\,r_3\,r_4\,+
\,\ell\,u_1\,u_3\,u_4\,+\,r\,u\,(\ell\,r\,+\,m\,u)\,+
\,4\,\ell\,m\,r\,u\,(s_1^2\,+\,s_3^2\,+\,s_4^2)\,\nn\\
&& +\,4\,\ell\,m\,(\ell\,r\,+\,m\,u)(s_{13}^2\,+\,s_{14}^2\,+
\,s_{34}^2)\,+\,16\,\ell^2\,m^2\,s_{134}^2\Big)\,\nn\\
&& + \,2\,(\ell\,r\,-\,m\,u)^2\,\Big(c_{134}\,s_{134}\,
(c_2^2\,+\,s_2^2)\,-\,c_2\,s_2\,(s_{13}^2\,+\,s_{14}^2\,+
\,s_{34}^2\,+\,2\,s_{134}^2)\Big)\Big ]\nn\\
\s_4\,&=&\,\fft{2}{W^2}\,\Big [c_4\,s_4\,\Big(m\,r_1\,r_2\,r_3\,+
\,\ell\,u_1\,u_2\,u_3\,+\,r\,u\,(\ell\,r\,+\,m\,u)\,+
\,4\,\ell\,m\,r\,u\,(s_1^2\,+\,s_2^2\,+\,s_3^2)\,\nn\\
&& +\,4\,\ell\,m\,(\ell\,r\,+\,m\,u)(s_{12}^2\,+\,s_{13}^2\,+
\,s_{23}^2)\,+\,16\,\ell^2\,m^2\,s_{123}^2\Big)\,\nn\\
&& +\,2\,(\ell\,r\,-\,m\,u)^2\,\Big(c_{123}\,s_{123}\,(c_4^2\,+
\,s_4^2)\,-\,c_4\,s_4\,(s_{12}^2\,+\,s_{13}^2\,+\,s_{23}^2\,+
\,2\,s_{123}^2)\Big)\Big ]\nn\\
B_{(1)}\,&=&\,\fft{2}{a\,\bar\rho}\,\Big[c_{1234}\,
\Big(\ell\,u\,(a^2\,+\,r^2)\,+\,\,m\,r(a^2\,-\,u^2)\Big )\,\nn\\
&& -\,s_{1234}\,\Big (\ell\,u_0\,(a^2+r^2)\,+\,m\,r_0\,
(a^2\,-\,u^2)\,+\,4\,\ell\,m\,(\ell\,r\,-\,m\,u) \Big)\Big]d\phi,\nn\\
A_{\1 1}\,&=&\,\fft{2\,c_1\,s_1}{a\,\bar\rho}\,[\,a^2\,(\ell\,r\,-
\,m\,u)-r\,u\,(m\,r_0\,+\,\ell\,u_0)\,]d\phi,\nn\\
\cA_\1^1\,&=&\,\fft{2\,c_3\,s_3}{a\,\bar\rho}\,[\,a^2\,(\ell\,r\,-
\,m\,u)-r\,u\,(m\,r_0\,+\,\ell\,u_0)\,]d\phi,\nn\\
A_{\1 2}&=&\fft{2}{a\,\bar\rho}\,\Big[s_2\,c_{134}\,
\Big(r\,u\,(\ell\,r-m\,u)+a^2\,(m\,r+\ell\,u)\Big)\nn\\
&&\qquad\qquad-c_2\,s_{134}\,\Big(r_0\,u_0\,(\ell\,r-m\,u)+a^2\,
(m\,r_0+\ell\,u_0)    \Big)      \Big]d\phi,\nn\\
\cA_\1^2&=&\fft{2}{a\,\bar\rho}\,\Big[s_4\,c_{123}\,
\Big(r\,u\,(\ell\,r-m\,u)+a^2\,(m\,r+\ell\,u)\Big)\nn\\
&&\qquad\qquad-c_4\,s_{123}\,\Big(r_0\,u_0\,(\ell\,r-m\,u)+a^2\,
(m\,r_0+\ell\,u_0)    \Big)      \Big]d\phi\,.\label{3dimexprs}
\eea

    Lifting back to four dimensions, the metric is given by
\bea
ds_4^2\,=\,-\fft{\bar\rho}{W}\,(dt\,+\,B_{(1)})^2\,+
\,W\,\Big(\fft{dr^2}{\D_r}\,+\,\fft{du^2}{\D_u}\,+
\,\fft{\D_r\,\D_u}{a^2\,\bar\rho}\,d\phi^2\Big)\,,
\eea
and the 4 four-dimensional gauge potentials are given in terms of the
three-dimensional expressions in (\ref{3dimexprs}) by
\bea
\hat A_{\1 1} &=& (A_{\1 1} + \sigma_1\, \cB_\1) + \sigma_1\, dt\,,\nn\\
\hat A_{\1 2} &=& (A_{\1 2} + \sigma_2\, \cB_\1) + \sigma_2\, dt\,,\nn\\
\hat \cA_{\1}^1 &=& (\cA_{\1}^1 + \sigma_3\, \cB_\1) + \sigma_3\, dt\,,\nn\\
\hat \cA_{\1}^2 &=& (\cA_{\1}^2 + \sigma_4\, \cB_\1) + \sigma_4\, dt\,.
\eea
The dilatons $(\varphi_1,\varphi_2,\varphi_3)$ and axions 
$(\chi_1,\chi_2,\chi_3)$ are simply given by their three-dimensional 
expressions in (\ref{3dimexprs}).

\section{Supersymmetry of Rotating AdS Black Holes}

   A complete discussion of the supersymmetry of the charged solutions
we have constructed is rather involved, and we shall not present it here.
There should be no essential difference between the supersymmetry
of the multi-charge solutions and the case where all charges are set
equal.  Thus for the purpose of discussions the fractions of supersymmetry
that are preserved, it suffices to consider just the case of charged
rotating black holes in gauged Einstein-Maxwell supergravity.  The results
are mostly known from previous literature, but it is perhaps useful
to collect them together here.

  Setting all the charges equal, in framework where the two 
magnetically-charge fields have been dualised to electrically-charged fields,
as in (\ref{d4lag4}) and (\ref{bfield}), our solution reduces to the standard
Kerr-Newman charged rotating black hole with a cosmological constant.  
This is given by
\bea
ds_4^2 &=& -\fft{\Delta_r}{\rho^2}\, (dt - a\, \sin^2\theta\,
d\phi)^2 + \rho^2 \, \Big( \fft{dr^2}{\Delta_r} +
\fft{d\theta^2}{\Delta_\theta} \Big) 
   + \fft{\Delta_\theta\, \sin^2\theta}{\rho^2}\, [a\, dt - (r^2 +
   a^2)d\phi]^2\,,\nn\\
A_\1  &=& \fft{2Q\, r}{\rho^2} [dt - a\, \sin^2\theta\, 
    d\phi]\,,
\eea
where
\be
\rho^2=r^2 +a^2\, \cos^2\theta\,,\quad 
\Delta_r =(r^2 + a^2)(1+g^2 \,r^2) -2m\,  r + 
Q^2 \,,\quad \Delta_\theta = 1 - g^2\, a^2\, \cos^2\theta\,.
\ee
This satisfies the equations of motion following from the Lagrangian
\be
{\cal L} = e\,(R - \ft14 F_\2^2 - 6 g^2)\,.\label{emlag1}
\ee

     Equation (\ref{emlag1}) is the bosonic sector of  
Einstein-Maxwell supergravity, for which the gravitino transformation rule
is $\delta\psi_\mu = D_\mu\, \ep$, where
\be
D_\mu = \nabla_\mu - \ft{\im}{2}\, g\, A_\mu + \ft{\im}{8}\, 
    F_{\nu\rho}\, \Gamma^{\nu\rho}\, \Gamma_\mu + \ft12 g\, 
\Gamma_\mu\,.
\ee
From this, we can easily calculate the integrability conditions
$M_{\mu\nu}\, \eta \equiv [D_\mu\, D_\nu]\, \eta=0$ for the existence of
a Killing spinor $\eta$.  The fraction of preserved supersymmetry is
determined by the number of common zero-eigenvalue eigenspinors of the
set of matrices $M_{\mu\nu}$.  We find the following:
\bigskip

\noindent\underline{{$g=0$, $a=0$}}:

   In this ungauged case, with no rotation, there are two common zero 
eigenvalues if $Q=m$, and two if $Q=-m$.  Thus a fraction $\ft12$ of
the supersymmetry is preserved.
\medskip

\noindent\underline{{$g=0$, $a\ne 0$}}: 

   In this ungauged case with rotation, there are again two common zero 
eigenvalues if $Q=m$ or $Q=-m$, and so again $\ft12$ supersymmetry is 
preserved.
\medskip

\noindent\underline{{$g\ne 0$, $a = 0$}}:

   In this gauged case without rotation, there are again two common
zero eigenvalues if $Q=m$ or $Q=-m$, and again $\ft12$ supersymmetry
is preserved.
\medskip

\noindent\underline{{$g\ne 0$, $a\ne 0$}}:

  In the gauged case with rotation, there is just one common zero eigenvalue,
arising in any of the four cases $m= \pm Q \pm a\, g\, Q$.  Thus four
any of these four sign choices, we find $\ft14$ of the supersymmetry 
is preserved.
\medskip

    It is worth remarking that our result in the case of $g\ne0$ and $a\ne 0$
(which also encompasses the previous three specialisations) is in complete
agreement with the discussion in Kostelecky and Perry \cite{kosper}.

\end{document}